\documentclass[usenatbib]{mnras}

\usepackage{newtxtext}
\usepackage[varvw]{newtxmath} 
\usepackage{graphicx}          
\usepackage{microtype}

\graphicspath{{./figures/}}
\hypersetup{linkcolor=red}     % red equations

\def\equationautorefname~#1\null{equation~(#1)\null}

\title[JWST Ultramassive Galaxy Sample -- I]{The JWST Ultramassive Galaxy Sample -- I. Overview, Photometry, and Stellar Kinematics: Selecting Optimal Template Subsets via Non-Negative Matrix Factorization}
    
\author[M. Cappellari, D. D. Nguyen, and S. A. Kassin]{
Michele Cappellari$^{1}$\thanks{E-mail: michele.cappellari@physics.ox.ac.uk},
Dieu D. Nguyen$^{2}$\thanks{E-mail: dieun@umich.edu} and Susan A. Kassin$^{3}$
\\
$^{1}$Sub-Department of Astrophysics, Department of Physics, University of Oxford, Denys Wilkinson Building, Keble Road, Oxford, OX1 3RH, UK\\
$^{2}$Department of Astronomy, University of Michigan, 1085 South University Avenue, Ann Arbor, MI 48109, USA\\
$^{3}$Space Telescope Science Institute, 3700 San Martin Drive, Baltimore, MD 21218, USA
}

\date{Submitted to MNRAS on 2026 September 18}

\pagerange{\pageref{firstpage}--\pageref{lastpage}} \pubyear{2026}

\begin{document}

\label{firstpage}
\maketitle

\begin{abstract}
We introduce the JWST Ultramassive Galaxy Sample, comprising 8 extreme early-type galaxies ($M_\star \gtrsim 2\times10^{12}\,\mathrm{M}_\odot$) drawn as a representative subset from a comprehensive full-sky census of over 100 such objects. This program aims to dynamically measure their central supermassive black holes (BHs) to probe the extreme high-mass end of BH--galaxy scaling relations, a critical regime where the physical mechanisms governing BH growth and host-galaxy co-evolution are expected to fundamentally change. In this first paper, we present the foundational data products: detailed photometric models and spatially resolved nuclear stellar kinematics. To accurately map the stellar mass distribution from the BH sphere of influence out to the extended halo, we combine diffraction-limited JWST/NIRCam imaging with wide-field DESI Legacy Surveys data, parametrizing the surface brightness using Multi-Gaussian Expansion (MGE). We extract stellar kinematics from JWST/NIRSpec integral-field observations using the penalized pixel-fitting (\textsc{pPXF}) method. To fully exploit the continuous $1.66-3.17\,\mu\mathrm{m}$ wavelength coverage---avoiding the telluric gaps of ground-based empirical libraries and the computational burden of massive theoretical grids---we introduce the use of near-separable Non-Negative Matrix Factorization (NMF). This algorithm identifies an optimal, highly compressed subset of synthetic spectra from the updated BOSZ library, perfectly capturing its spanning power at a fraction of the computational cost. The resulting kinematic maps reveal that these ultramassive systems are overwhelmingly slow rotators, frequently harboring kinematically decoupled cores. These high-quality photometric and kinematic data provide the inputs required for the robust Jeans Anisotropic Modelling (JAM) of the central BH masses in Paper II.
\end{abstract}

\begin{keywords}
galaxies: kinematics and dynamics -- galaxies: elliptical and lenticular, cD -- galaxies: structure -- galaxies: nuclei -- techniques: imaging spectroscopy -- methods: data analysis
\end{keywords}

\section{Introduction}
\label{sec:intro}

Supermassive black holes (BHs) are a fundamental component of massive galaxies. Their masses, $M_{\rm BH}$, tightly correlate with the large-scale properties of the host spheroid, most notably its stellar velocity dispersion, $\sigma_\star$ \citep{Ferrarese2000, Gebhardt2000bh}, and its total (or bulge) stellar mass, $M_\star$ \citep{Marconi2003, Haring2004}. These scaling relations are generally interpreted as evidence of a coupled assembly history between BHs and their host galaxies, driven by shared gas supplies, feedback from active galactic nuclei  \citep{Silk1998, DiMatteo2005}, and the hierarchical averaging produced by repeated mergers  \citep{Peng2007, Jahnke2011}. See review in \citet{Kormendy2013review}. Establishing the exact form and intrinsic scatter of these relations is therefore central to anchoring theoretical models of galaxy evolution. Yet, the high-mass extreme of these relations remains poorly constrained: the galaxies hosting the most massive BHs are exceedingly rare and typically found at large distances, making the angular size of the BH's gravitational sphere of influence difficult to resolve.

This poorly sampled regime is of particular physical interest because the most massive early-type galaxies are not merely scaled-up versions of lower-mass systems. Above $M_\star \sim 10^{12}\,\mathrm{M}_\odot$, the late-time assembly of galaxies is expected to be overwhelmingly dominated by dissipationless (dry) mergers \citep{Oser2010, Naab2017}. Observationally, these systems are commonly slow rotators \citep{Emsellem2011p3, Cappellari2013p20}, feature shallow, ``cored'' central surface-brightness profiles \citep{Faber1997, Lauer2005}, and frequently display complex, prolate-like rotation indicative of triaxiality and chaotic merger histories \citep{Krajnovic2018prolate, Krajnovic2026}. See reviews in \citet{Cappellari2016, Cappellari2026}. Crucially, dry mergers can substantially increase both $M_\star$ and $M_{\rm BH}$ while leaving $\sigma_\star$ comparatively unchanged \citep{Ciotti2007, Naab2009}. As a result, the highest-mass galaxies may undergo a structural transition, shifting from a regime where $M_{\rm BH}$ primarily tracks $\sigma_\star$ to one where it scales more directly with $M_\star$ \citep{Lauer2007, Krajnovic2018channels}. Measuring this potential transition would provide direct constraints on the assembly channels of the Universe's largest black holes and galaxies.

There are already tantalizing indications that such a transition occurs. Stellar-dynamical measurements in nearby brightest cluster galaxies have revealed ultramassive BHs with masses of order $10^{10}\,\mathrm{M}_\odot$, lying substantially above the extrapolation of the canonical $M_{\rm BH}$--$\sigma_\star$ relation \citep{McConnell2011, McConnell2012, Thomas2016, Mehrgan2019}. Strong gravitational lensing has also provided an independent detection of an ultramassive BH in the cluster Abell~1201 \citep{Nightingale2023}. However, the number of secure dynamical measurements in galaxies with $M_\star \gtrsim 10^{12}\,\mathrm{M}_\odot$ remains too sparse, and the data quality not always sufficient, to definitively distinguish a systematic change in the scaling relations from intrinsic scatter or sample selection biases. Volume-limited integral-field spectroscopic surveys, such as the MASSIVE survey \citep{Ma2014} and the MUSE Most Massive Galaxies survey \citep{Krajnovic2018prolate}, have pushed into this high-mass regime. Ground-based efforts nevertheless remained limited by the absence of a homogeneous all-sky sample and by the difficulty of resolving BH spheres of influence in the more distant members.

Robust stellar-dynamical BH measurements require spatially resolved kinematics that penetrate the sphere of influence, coupled with high-resolution photometry to accurately map the underlying stellar mass distribution. While ground-based adaptive optics (AO) have delivered important measurements, their performance in the shallow cores of giant early-type galaxies is severely limited by the lack of suitably bright guide stars, fluctuating seeing halos, telluric absorption, and large uncertainties in the point-spread function (PSF). The James Webb Space Telescope (JWST) eliminates these bottlenecks. The stable, diffraction-limited PSF of JWST, combined with the integral-field capabilities of NIRSpec \citep{Jakobsen2022,Boker2022} and matched near-infrared imaging from NIRCam \citep{Rieke2023}, provides an unprecedented platform for nuclear dynamics. The ability of such near-infrared observations to accurately constrain BH masses has been clearly demonstrated in recent simulations for ultramassive galaxies \citep{Nguyen2023, Nguyen2026z1-2} and validated by early JWST/NIRSpec dynamical measurements in nearby systems \citep{Nguyen2025, Nguyen2026maser, Nguyen2026m81}.

In this work, we present high-resolution JWST/NIRSpec integral-field spectroscopy and JWST/NIRCam imaging for a full-sky-selected sample of 8 ultramassive galaxies ($M_\star \gtrsim 2\times 10^{12}\,\mathrm{M}_\odot$) spanning a wide range of environments. Because robust dynamical modelling requires exquisite observational inputs, this paper focuses entirely on the foundational data analysis. We construct high-fidelity stellar luminosity models by seamlessly combining the high-resolution JWST imaging with wide-field ground-based data, and we extract the nuclear stellar kinematics from the NIRSpec observations using the penalized pixel-fitting method (\textsc{pPXF}; \citealt{Cappellari2004, Cappellari2017, Cappellari2023}). 

The photometric models and kinematic maps derived in this work provide the necessary ingredients to accurately map the gravitational potential of these extreme systems. In a companion paper, we will use these data products in combination with Jeans Anisotropic Modelling (JAM; \citealt{Cappellari2008, Cappellari2020, Cappellari2026jam}) to dynamically measure the central BH masses. Ultimately, this homogeneous dataset will allow us to test whether the most massive galaxies host BHs that systematically depart from the canonical scaling relations, and whether $M_\star$ supersedes $\sigma_\star$ as the primary predictor of BH mass in the ultramassive regime.

\section{Sample Selection}
\label{sec:sample}

To robustly measure the masses of ultramassive black holes and establish whether their scaling relations transition from a $\sigma_\star$-driven to an $M_\star$-driven regime, we require a systematically selected, volume-limited sample of the most massive galaxies in the Universe. Because such extreme objects are exceedingly rare, one must search across the entire sky and reach distances considerably larger than those probed by current volume-limited integral-field surveys such as ATLAS$^{\rm 3D}$ \citep{Cappellari2011p1} or MASSIVE \citep{Ma2014}. 

Our sample is a subset of the parent sample of 101 ultramassive galaxies presented in \citet{Nguyen2023}, where all details of the selection and sample properties are given. In brief, that parent sample was constructed by combining the full-sky, homogeneous near-infrared photometry of the 2MASS Redshift Survey \citep[2MRS;][]{Huchra2012} with the NASA/IPAC Extragalactic Database compilation of redshift-independent distances \citep[NED-D;][]{Steer2017}. Where redshift-independent distances were unavailable or poorly constrained, we derived angular-size distances, $D_{\rm A}$, and luminosity distances, $D_{\rm L}$, from the systemic redshifts, assuming a standard flat cosmology. The 2MASS $K_s$-band is largely insensitive to dust attenuation and offers a stellar mass-to-light ratio that varies within a factor of roughly two to three, making it an optimal proxy for robust stellar mass approximations. We derived total $K_s$-band absolute magnitudes, $M_K$, after applying foreground Galactic extinction corrections. 

To estimate the stellar masses of the parent sample, we utilized the tight empirical correlation between dynamical stellar mass and $K_s$-band luminosity, $\lg (M_\star) = 10.58 - 0.44 \times (M_K + 23)$, calibrated via detailed Jeans dynamical models of early-type galaxies from the ATLAS$^{\rm 3D}$ survey \citep[eq.~2]{Cappellari2013apjl}. We restricted our selection to the extreme high-mass end of the galaxy mass function by enforcing a lower limit of $M_\star \gtrsim 2 \times 10^{12}\,\mathrm{M}_\odot$.

The fundamental observational requirement for a reliable dynamical BH mass measurement is that the target's sphere of influence, $r_{\rm SOI} = G M_{\rm BH} / \sigma_\star^2$, must be at least marginally resolved by the telescope's point-spread function (PSF). We predicted the expected $M_{\rm BH}$ for each candidate using both the standard $M_{\rm BH}$--$\sigma_\star$ relation and the $M_{\rm BH}$--$M_\star$ relation \citep{Krajnovic2018channels}. For the stellar velocity dispersion, we adopted a characteristic value of $\sigma_\star \approx 300\,\mathrm{km\,s^{-1}}$, which is typical for nearby slow-rotator core galaxies of this mass and varies only weakly at the highest masses. 

We selected only those galaxies whose predicted $r_{\rm SOI}$ is larger than the exceptionally stable $0\farcs15$ full-width at half-maximum (FWHM) PSF of JWST/NIRSpec. Given the characteristic masses involved, this spatial resolution requirement effectively restricts our survey to galaxies at angular-size distances $D_{\rm A} < 500$\,Mpc.

The combination of the mass limit ($M_\star \gtrsim 2 \times 10^{12}\,\mathrm{M}_\odot$) and the spatial resolution constraint ($r_{\rm SOI} \ge 0\farcs1$) yields a rigorously defined, full-sky sample of exactly 8 ultramassive galaxies. Our selection surveys a comoving volume approximately 150 times larger than that of the MASSIVE survey, yet remains specifically tailored for the diffraction-limited capabilities of JWST. Crucially, by not imposing any prior on the local environment density, this sample encompasses diverse galactic environments ranging from relative isolation to the centres of dense galaxy clusters. This environmental diversity is essential to definitively separate generic high-mass deviations in the BH scaling relations from the specific evolutionary pathways of brightest cluster galaxies.

\section{Data}
\label{sec:data}

\subsection{Photometric data}
\label{sec:photometry}

\begin{table*}
\centering
\caption{JWST NIRCam photometric observations for the eight candidate ultramassive galaxies in our sample. The available calibrated FITS products use the \texttt{SUB640} subarray, \texttt{CLEAR} pupil, \texttt{BRIGHT1} readout pattern, and filters \texttt{F115W} and \texttt{F250M}. The exposure time listed is the effective exposure time recorded in the FITS headers.}
\label{tab:photometry}
\begin{tabular}{lccccc}
\hline
Target Name & RA (J2000) & Dec (J2000) & Filters & Total Exposure Time (s) & Date of Observation (UT) \\
\hline
2MASX J00034964+0203594 & 00 03 49.67 & +02 03 59.4 & \texttt{F115W}, \texttt{F250M} & -- & Not yet available \\
2MASX J01313288+0033210 & 01 31 32.92 & +00 33 21.5 & \texttt{F115W}, \texttt{F250M} & 3817 & 2026 Jul 10 \\
2MASX J04385250-2206391 & 04 38 52.53 & --22 06 39.1 & \texttt{F115W}, \texttt{F250M} & -- & Not yet available \\
2MASX J10194430-0038173 & 10 19 44.27 & --00 38 18.0 & \texttt{F115W}, \texttt{F250M} & -- & Not yet available \\
2MASX J12052321+1022461 & 12 05 23.23 & +10 22 46.2 & \texttt{F115W}, \texttt{F250M} & 3817 & 2026 Jun 17 \\
2MASX J15165808-0106394 & 15 16 58.08 & --01 06 38.9 & \texttt{F115W}, \texttt{F250M} & 3817 & 2026 Aug 9 \\
2MASX J15342642-0119003 & 15 34 26.43 & --01 19 00.2 & \texttt{F115W}, \texttt{F250M} & 3817 & 2026 Aug 13 \\
2MASX J22354078+0129053 & 22 35 40.80 & +01 29 05.6 & \texttt{F115W}, \texttt{F250M} & 3817 & 2026 May 29 \\
\hline
\end{tabular}
\end{table*}

Accurate stellar-dynamical measurements of supermassive black hole masses require a precise determination of the host galaxy's stellar mass distribution. In particular, we must accurately trace the stellar surface brightness profile from the scales of the black hole sphere of influence ($r_\mathrm{SOI} \approx 0\farcs1 - 0\farcs3$) out to the extended galaxy halo. To achieve this, we combined ultra-high-resolution space-based imaging from JWST with wide-field ground-based imaging. 

\subsubsection{JWST NIRCam Photometry}

We acquired deep, high-resolution near-infrared imaging for our complete sample of 8 ultramassive galaxies using the Near-Infrared Camera (NIRCam) on board JWST (GO Program 8217; Co-PIs: D. Nguyen and M. Cappellari). The observations were designed to constrain the nuclear stellar mass profile precisely in the same photometric bands that are closest to our kinematics, thereby minimising uncertainties associated with dust extinction and radial stellar population gradients.

We observed simultaneously in the short-wavelength (SW) and long-wavelength (LW) channels using the F115W ($1.15\,\mu\mathrm{m}$) and F250M ($2.50\,\mu\mathrm{m}$) filters, respectively. The F250M filter was specifically chosen to trace the bulk of the stellar mass (as it closely matches the $K_s$ band used for our sample selection), while the combination with F115W provides critical colour information to constrain any central variations in the stellar mass-to-light ratio ($M_\star/L$). 

To optimise observing efficiency and avoid inter-chip gaps, we configured NIRCam using the \texttt{SUB640} subarray on Module B. This provides a field of view of roughly $40\arcsec \times 40\arcsec$ in the LW channel (pixel scale $0\farcs063\,\mathrm{pixel}^{-1}$), comfortably encompassing the central effective radius of our targets without the need for large mosaics. We employed the \texttt{BRIGHT1} readout pattern with 10 groups per integration and 6 integrations per exposure. We used the \texttt{SMALL-GRID-DITHER} pattern with 8 sub-pixel positions to optimally sample the point spread function (PSF) and mitigate detector artifacts and bad pixels. This setup yielded a total planned on-target exposure time of $4019.389\,\mathrm{s}$ per galaxy for both filters simultaneously. A summary of the JWST NIRCam observations is provided in \autoref{tab:photometry}.

\subsubsection{DESI Legacy Surveys DR10 Photometry}

Because the JWST NIRCam \texttt{SUB640} footprint is restricted to the central $\approx 40\arcsec$, it truncates the outer envelopes of our giant elliptical galaxies, whose effective radii extend over tens of kiloparsecs. To properly normalise the total stellar mass and robustly constrain the outer boundary conditions for our dynamical models, wide-field imaging is mandatory.

We augmented the JWST data with optical images extracted from Data Release 10 (DR10) of the Dark Energy Spectroscopic Instrument (DESI) Legacy Imaging Surveys \citep{Dey2019}. The DESI Legacy Surveys provide deep, homogeneous, sky-subtracted optical imaging that covers our entire target sample. We downloaded $1024 \times 1024$ pixel cutouts in the $z$-band, which most closely bridges the optical and near-infrared regimes. With the native pixel scale of $0\farcs262\,\mathrm{pixel}^{-1}$, these cutouts cover a massive footprint of $\approx 4\farcm5 \times 4\farcm5$, guaranteeing that the sky background and the extended outer stellar haloes are properly sampled. These ground-based images were strictly used to anchor the Multi-Gaussian Expansion (MGE) models at large radii, while the high-resolution JWST data dictated the critical central mass distributions.

\subsection{Spectroscopic data}
\label{sec:spectroscopy}

\begin{table*}
\centering
\caption{JWST NIRSpec integral-field spectroscopic observations for the eight ultramassive galaxies in our sample. The FITS headers indicate the \texttt{G235H} grating, \texttt{F170LP} filter, \texttt{NRSIRS2RAPID} readout pattern, and four exposures for every target. The wavelength range and exposure times are taken from the reduced FITS headers.}
\label{tab:spectroscopy}
\begin{tabular}{lccccc}
\hline
Target Name & Wavelength Range & Spectral Res. ($R$) & Total Exposures & Total Exposure Time (s) & Date of Observation (UT) \\
\hline
2MASX J00034964+0203594 & $1.66$--$3.17\,\mu\mathrm{m}$ & $\sim2700$ & 4 & 8286 & 2026 Jul 20 \\
2MASX J01313288+0033210 & $1.66$--$3.17\,\mu\mathrm{m}$ & $\sim2700$ & 4 & 8286 & 2026 Aug 16 \\
2MASX J04385250-2206391 & $1.66$--$3.17\,\mu\mathrm{m}$ & $\sim2700$ & 4 & 8286 & 2025 Dec 18 \\
2MASX J10194430-0038173 & $1.66$--$3.17\,\mu\mathrm{m}$ & $\sim2700$ & 4 & 8286 & 2026 May 15 \\
2MASX J12052321+1022461 & $1.66$--$3.17\,\mu\mathrm{m}$ & $\sim2700$ & 4 & 8286 & 2026 Jan 28 \\
2MASX J15165808-0106394 & $1.66$--$3.17\,\mu\mathrm{m}$ & $\sim2700$ & 4 & 8286 & 2026 Jul 20 \\
2MASX J15342642-0119003 & $1.66$--$3.17\,\mu\mathrm{m}$ & $\sim2700$ & 4 & 8286 & 2026 Jul 16 \\
2MASX J22354078+0129053 & $1.66$--$3.17\,\mu\mathrm{m}$ & $\sim2700$ & 4 & 8286 & 2025 Nov 5 \\
\hline
\end{tabular}
\end{table*}

To spatially resolve the stellar kinematics within the gravitational sphere of influence of the central black holes, we obtained two-dimensional near-infrared integral-field spectroscopy using the JWST Near-Infrared Spectrograph (NIRSpec; \citealt{Jakobsen2022}) in IFS mode \citep{Boker2022}. Ground-based adaptive optics (AO) observations of such distant, core-profile galaxies are heavily compromised by the lack of bright natural guide stars and atmospheric seeing halos. In contrast, the stable, diffraction-limited point spread function (PSF) of JWST in space allows us to probe the critical central $0\farcs1 - 0\farcs3$ undisturbed.

We observed all 8 ultramassive galaxies using the high-resolution \texttt{G235H} grating combined with the \texttt{F170LP} filter. This configuration yields a spectral resolving power of $R \approx 2700$, which corresponds to an instrumental velocity dispersion of $\sigma_\mathrm{instr} \approx 47\,\mathrm{km\,s^{-1}}$. This instrumental resolution is optimally matched to our targets, whose intrinsic stellar velocity dispersions ($\sigma_\star \gtrsim 300\,\mathrm{km\,s^{-1}}$) easily dominate the instrumental broadening, ensuring precise extraction of the higher-order Gauss-Hermite moments ($h_3, h_4$). 

The \texttt{G235H/F170LP} setup provides contiguous wavelength coverage from $1.66$ to $3.17\,\mu\mathrm{m}$. This range is exceptionally rich in diagnostic stellar absorption features, crucially covering the prominent $^{12}$CO vibrational bandheads longward of $2.29\,\mu\mathrm{m}$. These CO features are among the most robust tracers of the stellar kinematics for old stellar populations in the near-infrared, suffering minimally from template-mismatch issues.

The NIRSpec IFS provides a $3\arcsec \times 3\arcsec$ field of view consisting of $0\farcs1$ spaxels. At the distances of our sample ($D_\mathrm{A} \approx 150 - 500\,\mathrm{Mpc}$), this field covers a physical radius of $\sim 1.5 - 3.5\,\mathrm{kpc}$, enveloping the black hole sphere of influence by a factor of at least $3$ to $5$. This coverage is the exact requirement for dynamically constraining the central black hole mass using the Jeans Anisotropic Models (JAM), as the kinematics seamlessly transition from the Keplerian regime dominated by the black hole out to the galaxy-dominated inner halo.

To mitigate detector artefacts, bad pixels, and cosmic rays, we executed a \texttt{4-POINT-DITHER} pattern. We employed the \texttt{NRSIRS2RAPID} readout pattern, which is optimized for reducing read noise in deep exposures, selecting 71 groups per integration and 1 integration per exposure. This observational strategy yielded a total planned on-target exposure time of $4201.6\,\mathrm{s}$ ($\approx 1.17\,\mathrm{hours}$) per galaxy. A summary of these spectroscopic observations is provided in \autoref{tab:spectroscopy}.

The NIRSpec IFU spatially undersamples the point-spread function (PSF), which lies close to the diffraction limit. For a compact source this means that the dispersed spectral trace lands on the detector at a sub-pixel phase that drifts with wavelength, so the trace crosses pixel boundaries at slightly different positions from one wavelength to the next. When such undersampled data are rectified into a three-dimensional cube with the standard 3D-drizzle algorithm, this wavelength-dependent phase drift turns into slow, periodic modulations of the spectra, the so-called ``wiggles'' \citep{Law2023}, which reach amplitudes of $6$--$9\%$ per spaxel close to unresolved point sources.

Rather than removing these features after the fact from extracted spectra \citep{Perna2023, Dumont2025, Shajib2025}, the Adaptive Trace Modeling (ATM) approach of \citet{Law2026} tackles them before the cube is assembled. It fits cubic B-spline models to the dispersed traces directly in native detector coordinates, drawing on the $\pm50$ adjacent columns to sample the full range of pixel phases, and then uses those models to oversample the detector frame onto a finer grid ahead of cube resampling. Emission that departs from the assumed smooth spline profile, such as spatially extended, scene-filling lines, is absorbed by an additional residual term. For our G235H data, an oversampling factor of $N=3$ brought the per-spaxel wiggle amplitude down to the $\sim1\%$ noise level while conserving the absolute flux to within $0.3\%$, in line with what \citet{Law2026} report for the G140H, G235M, and G395H modes.

This correction is available as the optional \texttt{adaptive\_trace\_model} step in Build 12.3 of the JWST calibration pipeline, and it is now recommended for IFU programs that rely on trustworthy spaxel-by-spaxel spectroscopy, as our does. We therefore applied it to our G235H/F170LP cubes, and in the same reduction resampled the spatial scale from the default $0\farcs1$ to $0\farcs05$.

\section{Multi-Gaussian Expansion Photometric Models}
\label{sec:mge}

\begin{figure*}
    \centering
    \includegraphics[width=\textwidth]{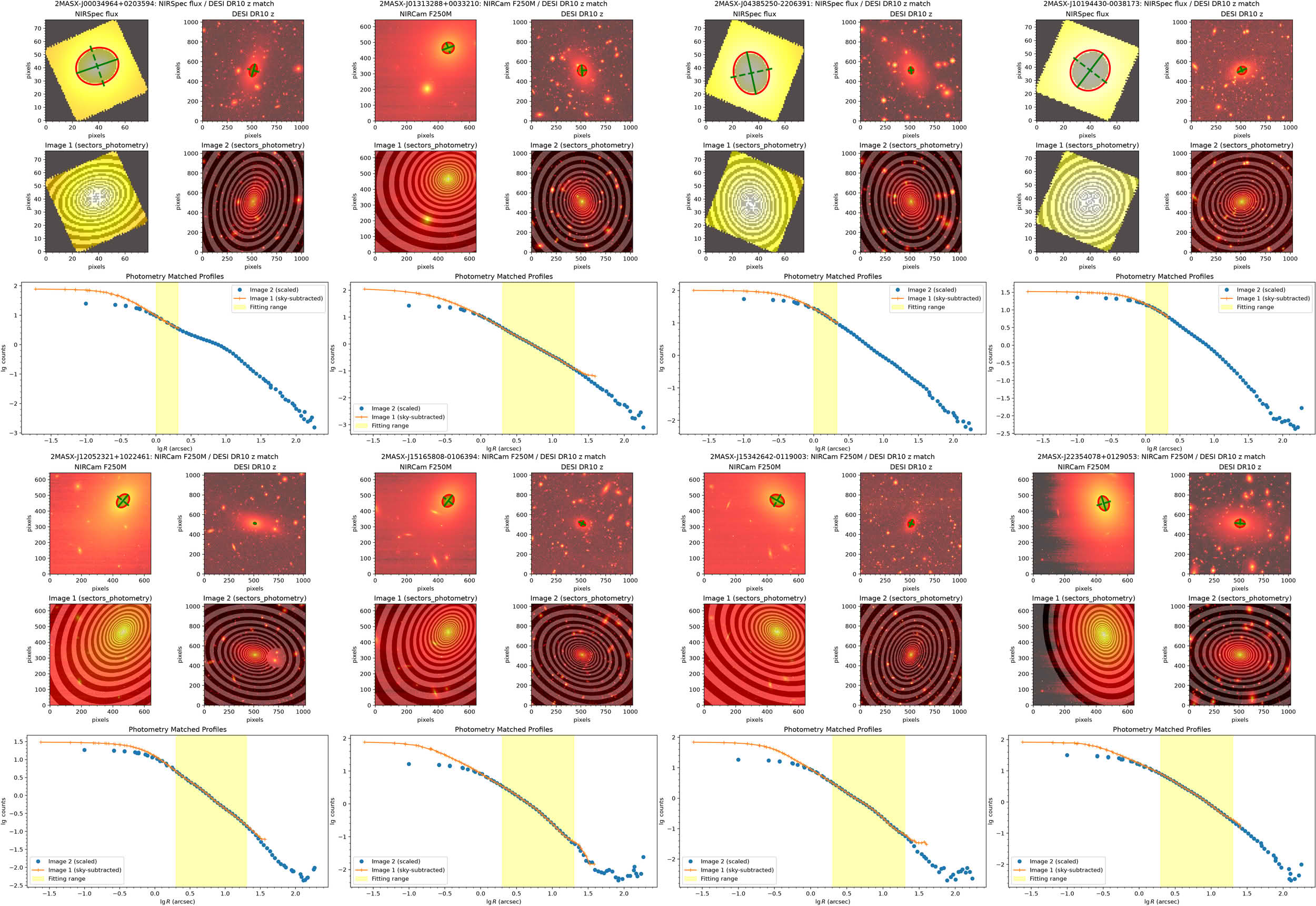}
    \caption{Diagnostic plots illustrating the photometric matching between the central high-resolution JWST data and the extended ground-based DESI imaging for eight of our ultramassive galaxies. For each galaxy, the procedure is detailed across three rows. \textbf{Top row:} The determination of the photometric centre, position angle, and global ellipticity via image moments for the JWST reference image (left; NIRCam F250M or collapsed NIRSpec flux) and the DESI Legacy Surveys DR10 $z$-band image (right). \textbf{Middle row:} The elliptical sectors (red contours) along which the radial surface brightness photometry was extracted for both datasets. \textbf{Bottom row:} The extracted 1D photometric profiles (log counts versus log radius). The yellow shaded region indicates the overlapping radial annulus used to match the two datasets. This optimal match is achieved using the \texttt{mge.PhotometryMatch} algorithm, which simultaneously fits for the residual sky background to be subtracted from the JWST image and the multiplicative flux scaling factor required to align the ground-based DESI image, which was independently sky-subtracted prior to the matching.}
    \label{fig:mge_profiles}
\end{figure*}

\begin{figure*}
    \centering
    \includegraphics[width=\textwidth]{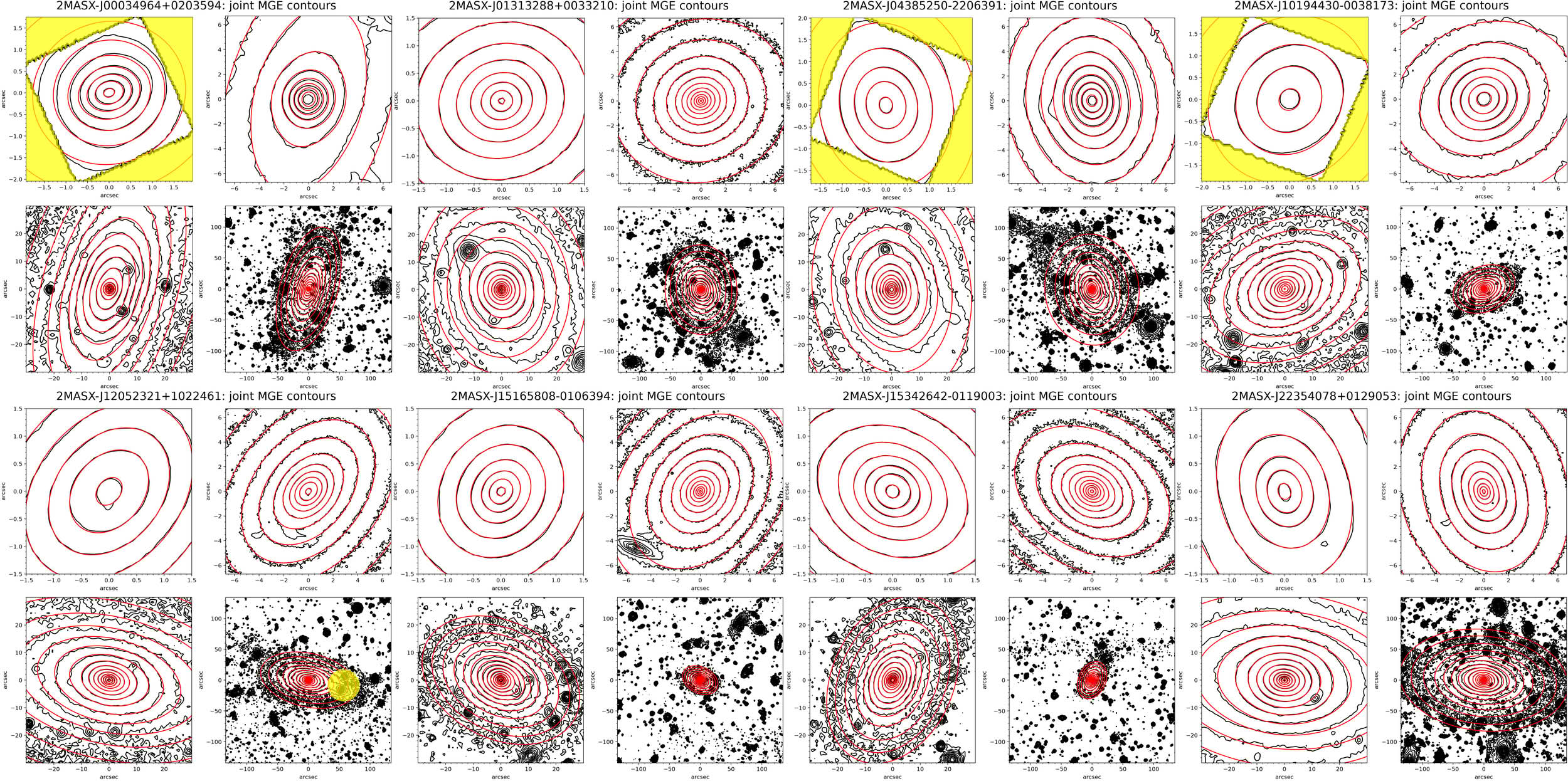}
    \caption{Joint Multi-Gaussian Expansion (MGE) photometric models for eight of the ultramassive galaxies in our sample, combining high-resolution JWST and wide-field DESI imaging. For each galaxy, the contours are displayed in a $2 \times 2$ grid of panels representing four logarithmically spaced fields of view, zooming out from $3\arcsec \times 3\arcsec$ (approximating the NIRSpec IFS footprint) to $300\arcsec \times 300\arcsec$. In each $2 \times 2$ block, the top two panels show the central regions observed with JWST, while the bottom two panels display the extended stellar halo captured by the DESI Legacy Surveys DR10 $z$-band images. The black contours trace the observed isophotes of the data. The red contours represent a single, global MGE model fitted simultaneously to the combined 1D photometric profiles of both datasets (see \autoref{sec:mge}). The MGE models are mathematically projected onto the 2D plane and rigorously convolved with the appropriate instrumental PSF for each panel. Panels highlighted with a yellow background indicate targets for which the NIRCam imaging is currently pending; for these galaxies, the central high-resolution contours are derived directly from the collapsed JWST NIRSpec IFS flux cubes.}
    \label{fig:mge_contours}
\end{figure*}

\begin{figure*}
    \centering
    \includegraphics[width=\textwidth]{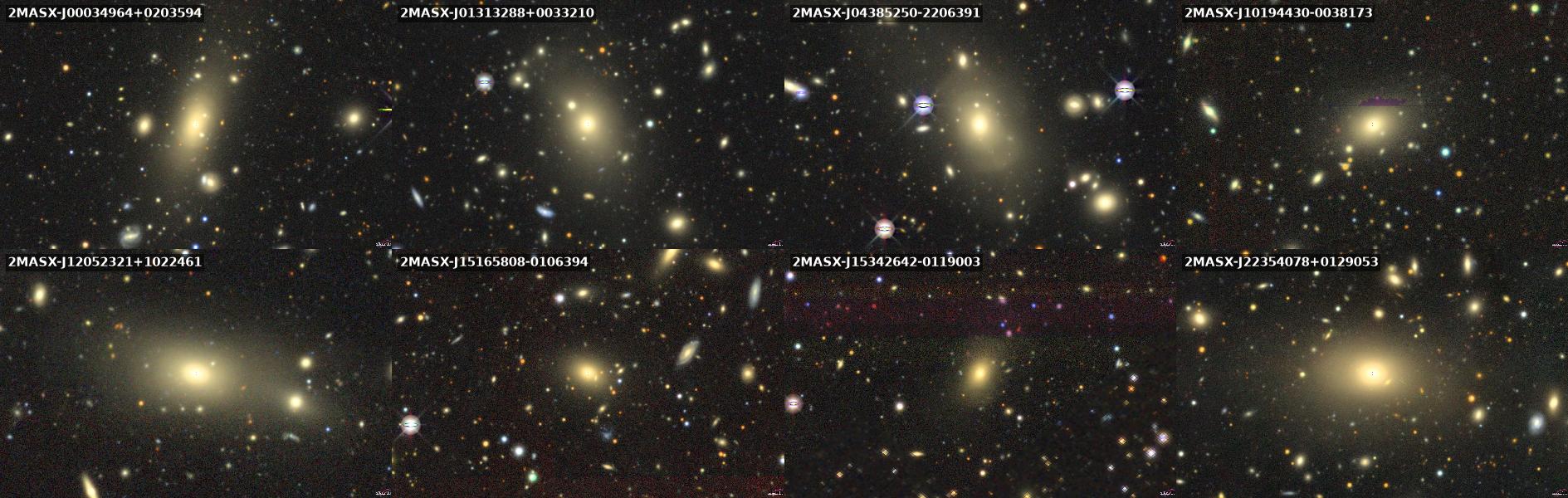} 
    \caption{Wide-field RGB composite images from the DESI Legacy Surveys DR10 for eight of the ultramassive galaxies in our sample, arranged in a $2 \times 4$ grid. Each cutout spans approximately $13\farcm8 \times 8\farcm8$ ($3156 \times 2004$ pixels at the native DESI scale of $0\farcs262\,\mathrm{pixel}^{-1}$), providing a comprehensive view of the diverse large-scale environments surrounding our targets. These environments range from relative isolation (e.g., 2MASX-J22354078+0129053) to the dense cores of massive galaxy clusters (e.g., 2MASX-J00034964+0203594). These deep, ground-based observations are crucial for accurately tracing the extended stellar envelopes and intra-cluster light of these giant ellipticals, thereby firmly anchoring the outer boundary conditions of our Multi-Gaussian Expansion (MGE) photometric models. The target identifiers are overlaid in the top-left corner of each panel. The images are constructed using the standard DESI colour mapping, with the $g$, $r$, and $z$ bands assigned to the blue, green, and red channels, respectively.}
    \label{fig:desi_rgb}
\end{figure*}

To derive the black hole masses from the observed stellar kinematics, our Jeans Anisotropic Models (JAM) require an accurate, deprojected 3D stellar mass distribution for each galaxy. We parametrized the stellar surface brightness profiles using the Multi-Gaussian Expansion (MGE) method \citep{Emsellem1994,Cappellari2002mge}, which represents the galaxy surface brightness as a sum of concentric, two-dimensional Gaussian functions. We performed the fitting using the Python package \textsc{mgefit}\footnote{Available at \url{https://pypi.org/project/mgefit/}} \citep{Cappellari2002mge}.

\subsection{Combining High-Resolution and Wide-Field Imaging}

A fundamental challenge in modelling the most massive early-type galaxies is that their extreme sizes (with effective radii extending over tens of kiloparsecs) far exceed the field of view of high-resolution space-based instruments, while ground-based surveys lack the requisite spatial resolution to probe the black hole sphere of influence. To overcome this, we combined the central high-resolution JWST images (NIRCam F250M, a $K$-band equivalent, for five galaxies, and collapsed NIRSpec flux cubes for the three targets currently lacking NIRCam coverage) with wide-field ground-based DESI DR10 $z$-band cutouts. By combining these different filters, we essentially assume that radial colour gradients are negligible---a highly reliable assumption for these old, passive galaxies. In any case, any minor colour variations would have a completely negligible effect on our dynamical models, which are overwhelmingly constrained by the distribution of the dynamical tracer light within the compact NIRSpec field of view. Furthermore, this approach bypasses any need to reconcile the differing photometric zeropoints between the two bands. As is standard practice in such combined analyses, we utilized the central, high-resolution JWST photometry as the absolute reference and simply scaled the outer, ground-based DESI profile to match it in the overlap region, ensuring a seamless structural transition without relying on absolute flux calibrations, which would not be equally accurate.

The procedure for each galaxy was strictly strictly data-driven and executed as follows:
\begin{enumerate}
    \item We first determined the photometric centre, global position angle, and characteristic ellipticity for both the JWST and DESI images independently using the \texttt{mge.find\_galaxy} routine. A common ellipticity was adopted to define the sectors for the subsequent radial profile extraction.
    \item We robustly estimated the sky background in the wide-field DESI images using \texttt{mge.sky\_level}, overriding the values manually only in cases of severe field crowding or excessively extended galaxy envelopes.
    \item To seamlessly stitch the high-resolution core to the wide-field halo, we used the \texttt{mge.PhotometryMatch} algorithm within the \textsc{MgeFit} package. This routine identifies an annular overlapping region (typically between $2\arcsec$ and $20\arcsec$) and solves simultaneously for the optimal flux scaling factor to convert the DESI counts to the JWST flux system, and the residual sky background level in the JWST images.
    \item We extracted the surface brightness along angular sectors using \texttt{mge.sectors\_photometry}. The JWST profile was used for the inner regions ($R \le 20\arcsec$), while the scaled DESI profile provided the boundary conditions for the outer halo.
    \item The joint 1D profiles were fitted simultaneously using the regularized MGE fitting routine \texttt{mge.fit\_sectors\_regularized}. To account for instrumental blurring, the MGE optimization rigorously convolved the model with the respective point spread functions (PSFs): a single Gaussian component for the JWST core (FWHM $\approx 0\farcs085$ for NIRCam, $0\farcs15$ for NIRSpec), and a double-Gaussian component for the DESI halo to accurately capture both the narrow core and broader seeing wings (FWHM of $0\farcs7$ and $1\farcs4$).
\end{enumerate}

\subsection{Photometric Fit Results}

The MGE method successfully reproduces the complex photometry of our ultramassive galaxy sample across several orders of magnitude in spatial scale and surface brightness. We graphically summarize the results of our photometric modelling in three composite mosaic figures:

\begin{itemize}
    \item \textbf{Matched Photometric Profiles:} \autoref{fig:mge_profiles} shows the combined 1D surface brightness profiles extracted along the major axis for all 8 galaxies. The plots highlight the excellent agreement in the overlap region between the scaled DESI data (outer points) and the sky-subtracted JWST data (inner points), demonstrating the robustness of our \texttt{mge.PhotometryMatch} scaling.
    \item \textbf{MGE Contour Maps:} To verify that our 1D regularized fits accurately capture the 2D morphology of the galaxies and their environments, \autoref{fig:mge_contours} presents a set of four contour panels for each galaxy. These panels span logarithmically spaced spatial scales, zooming sequentially from the central few arcseconds resolved by JWST out to the wide-field DESI footprint. The MGE models (red contours) perfectly track the observed isophotes (black contours) across all scales.
    \item \textbf{Large-scale Environments:} Finally, \autoref{fig:desi_rgb} displays wide-field RGB composite images generated from the DESI Legacy Surveys for the entire sample, providing context for the galactic environments, varying from relative isolation to the cores of dense clusters, from which the outer MGE boundary conditions are derived.
\end{itemize}

\section{Optimal Selection of Spectral Libraries for Kinematic Extraction}
\label{sec:nmf_selection}

Before proceeding to the dynamical analysis of our JWST observations, we make a brief but important methodological parenthesis. Extracting accurate stellar kinematics from high-quality spectra requires templates that can perfectly match the observed galaxy spectrum. In this section, we introduce a mathematically optimal approach for compressing massive spectral libraries into highly efficient, non-redundant subsets.

\subsection{The Separable Non-Negative Matrix Factorization Approach}
\label{sec:nmf_method}

In full-spectrum fitting---such as the approach implemented in the widely used software \textsc{pPXF}---a galaxy spectrum is modeled as a linear combination of template spectra, convolved with a line-of-sight velocity distribution and multiplied by smoothly varying polynomials. 

When the scientific objective is to recover star formation histories or detailed chemical abundances, it is generally necessary to use Single Stellar Population (SSP) models uniformly sampled across a grid of physical parameters (e.g., age, metallicity, and $\alpha$-enhancement). However, when the sole objective is to extract robust stellar kinematics, empirical or synthetic libraries of individual stars are often superior to SSPs. Individual stars provide the fitting algorithm with maximum freedom to reproduce the exact line-strength ratios and spectral morphology of the galaxy, without being constrained by the rigid assumptions of stellar evolution tracks and initial mass functions inherent to SSP models.

The primary drawback of using individual stellar libraries is their sheer size, which can encompass hundreds of thousands of spectra. Such libraries contain massive redundancies. Because the computational cost of the kinematic extraction scales with the number of templates, it is crucial to identify a small, computationally efficient subset (e.g., a few hundred stars) that retains the full spanning power of the original library. Previous approaches to this compression problem include uniform decimation across the physical parameter grid, or hierarchical clustering techniques, such as those successfully employed by the MaNGA Data Analysis Pipeline \citep{Westfall2019}.

Here, we present a method that is mathematically optimal for the specific mechanics of full-spectrum fitting. Consider the spectral library as a large matrix $\mathbf{L}$, where each column is an individual stellar spectrum. \textsc{pPXF} seeks to reconstruct the galaxy spectrum using a non-negative linear combination of a subset of these columns. Therefore, the ideal subset of, say, 100 stars is the specific set of 100 columns of $\mathbf{L}$ that can reconstruct \emph{every other column} in $\mathbf{L}$ with the minimum possible error, using strictly non-negative weights.

This formulation is exactly the problem that Non-Negative Matrix Factorization (NMF; \citealt{Lee1999}) was designed to solve. NMF has gained immense popularity in recent years within the machine learning community, particularly for tasks like facial recognition and image classification, because the non-negativity constraint forces the algorithm to learn an additive, ``parts-based'' representation of the data, rather than the abstract, often unphysical (e.g., negative flux) principal components produced by Singular Value Decomposition (SVD) or Principal Component Analysis (PCA).

Within the astrophysical literature, generalized NMF has previously been employed for spectral dimensionality reduction, most notably by \citet{Blanton2007} to derive the \texttt{kcorrect} galaxy templates. Their method elegantly used an empirical training set of observed galaxies to combine physical stellar population models into a few new, composite templates---an approach perfectly suited for modeling broadband photometry and low-resolution spectra. Our objective, however, requires a different formulation. First, mixing stellar models into composite spectra can subtly wash out exact atomic and molecular line depths, degrading the precision of higher-order kinematic moments; we instead require our templates to remain actual, unaltered individual stellar spectra. Second, our compression operates entirely on the theoretical library itself, without reference to an empirical training set of galaxies. This preserves the full spanning power of the parameter space, while the exact determination of which stellar types populate a given galaxy is best left to the native Non-Negative Least Squares (NNLS) optimization natively performed by \textsc{pPXF} during the final fit.

The specific requirement to extract exact, discrete columns from the library itself corresponds precisely to a distinct subclass of NMF known as Near-Separable Non-Negative Matrix Factorization. While the general NMF problem is non-convex and computationally NP-hard, the imposition of the near-separability condition---the strict mathematical assumption that the optimal basis vectors are actual, existing data points within the dataset itself---transforms it into a problem that can be solved in polynomial time. As described in the comprehensive monograph on NMF by \citet{Gillis2020}, several highly efficient algorithms have recently emerged to solve this specific factorization. For our purposes, we adopted the Successive Nonnegative Projection Algorithm (SNPA; \citealt{Gillis2014}; Algorithm 7.3 in \citealt{Gillis2020}), which currently represents the state-of-the-art for robustness and speed in near-separable NMF problems. 

By applying SNPA, we are guaranteed to select the most linearly independent, non-redundant set of actual stellar spectra capable of non-negatively reconstructing the entire parameter space of the parent library, without ever artificially blending their intrinsic spectral features. In the following subsection, we will apply this technique to the massive BOSZ grid of synthetic stellar spectra to construct our optimal kinematic template library.

\subsection{Optimal Template Selection from the BOSZ Synthetic Library}
\label{sec:nmf_application}

\begin{figure}
    \centering
    \includegraphics[width=\columnwidth]{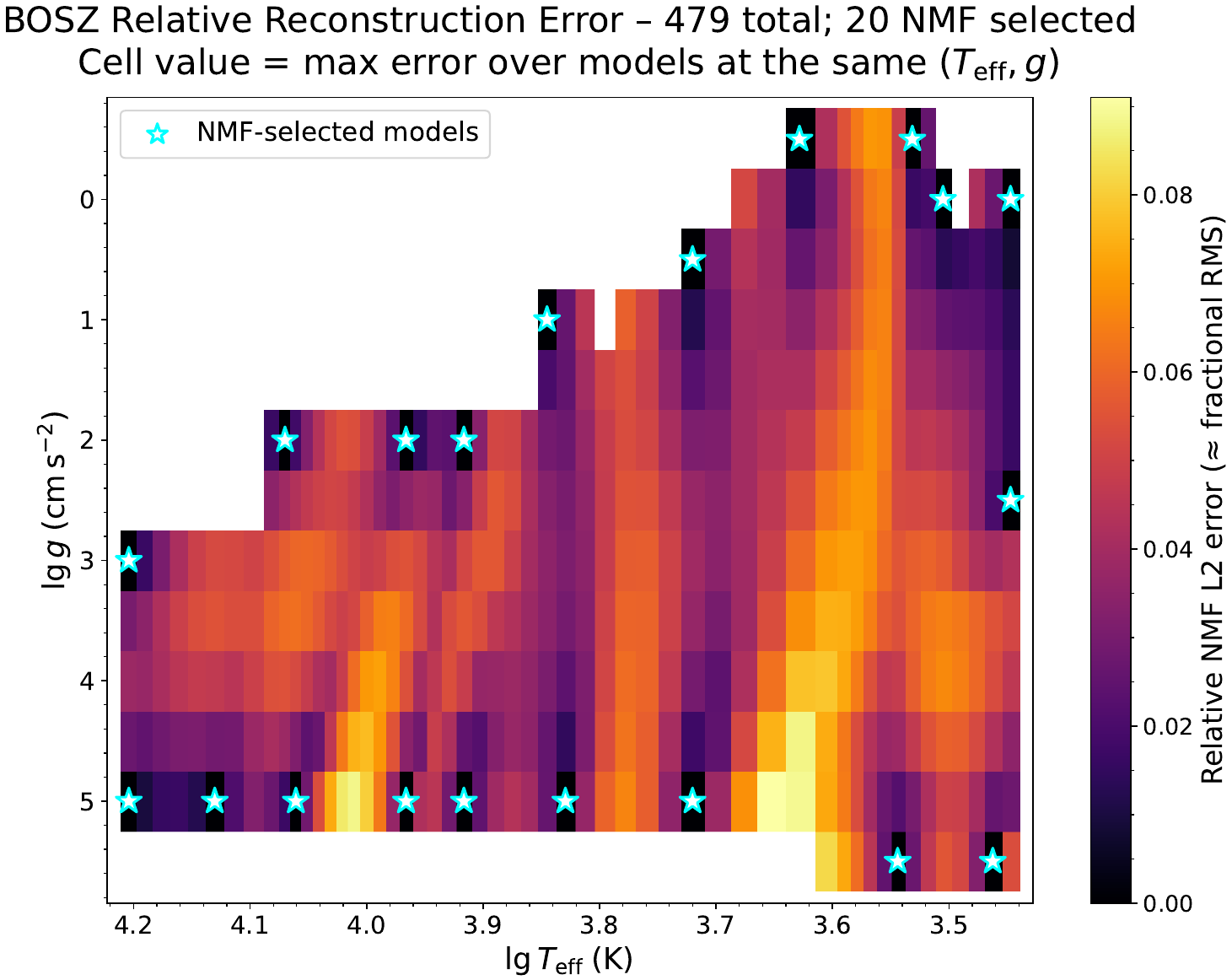}
    \caption{\textbf{Demonstration of the SNPA algorithm on a restricted BOSZ parameter space.} We restricted the BOSZ synthetic library to strictly solar metallicity ($[\mathrm{M/H}]=0$) and solar $\alpha$-abundance ($[\alpha/\mathrm{M}]=0$), and used the successive non-negative projection algorithm (SNPA) to select 20 representative models. The single panel shows the full set of BOSZ models in the $(\lg g, \lg T_\mathrm{eff})$ plane. Each model is represented by a colored rectangle whose color indicates its relative RMS reconstruction error when the spectrum is fitted with non-negative combinations of the 20 SNPA-selected models; the cyan stars mark the selected models. By construction, the error is zero at the location of each selected spectrum. The algorithm is given no information about the physical parameters: the template matrix is merely a continuous set of spectra. Nevertheless, it accurately recognizes the boundaries of the physical parameter space from spectral features alone: the selected templates (star symbols in the plot) not only identify the edges of the parameter ranges, but even the specific corners of the jagged distribution. Spectra along the top and bottom edges are generally sufficient to produce nearly constant RMS residuals at interior points, giving rise to vertical bands of nearly constant color connecting pairs of selected BOSZ models. This provides an intuitive picture of the NMF idea, which differs fundamentally from the traditional strategy of uniformly sampling the entire interior of parameter space. The preferential selection of boundary and corner models, together with the low reconstruction errors across the grid, validates this approach.}
    \label{fig:nmf_demo}
\end{figure}

\begin{figure*}
    \centering
    \includegraphics[width=.49\textwidth]{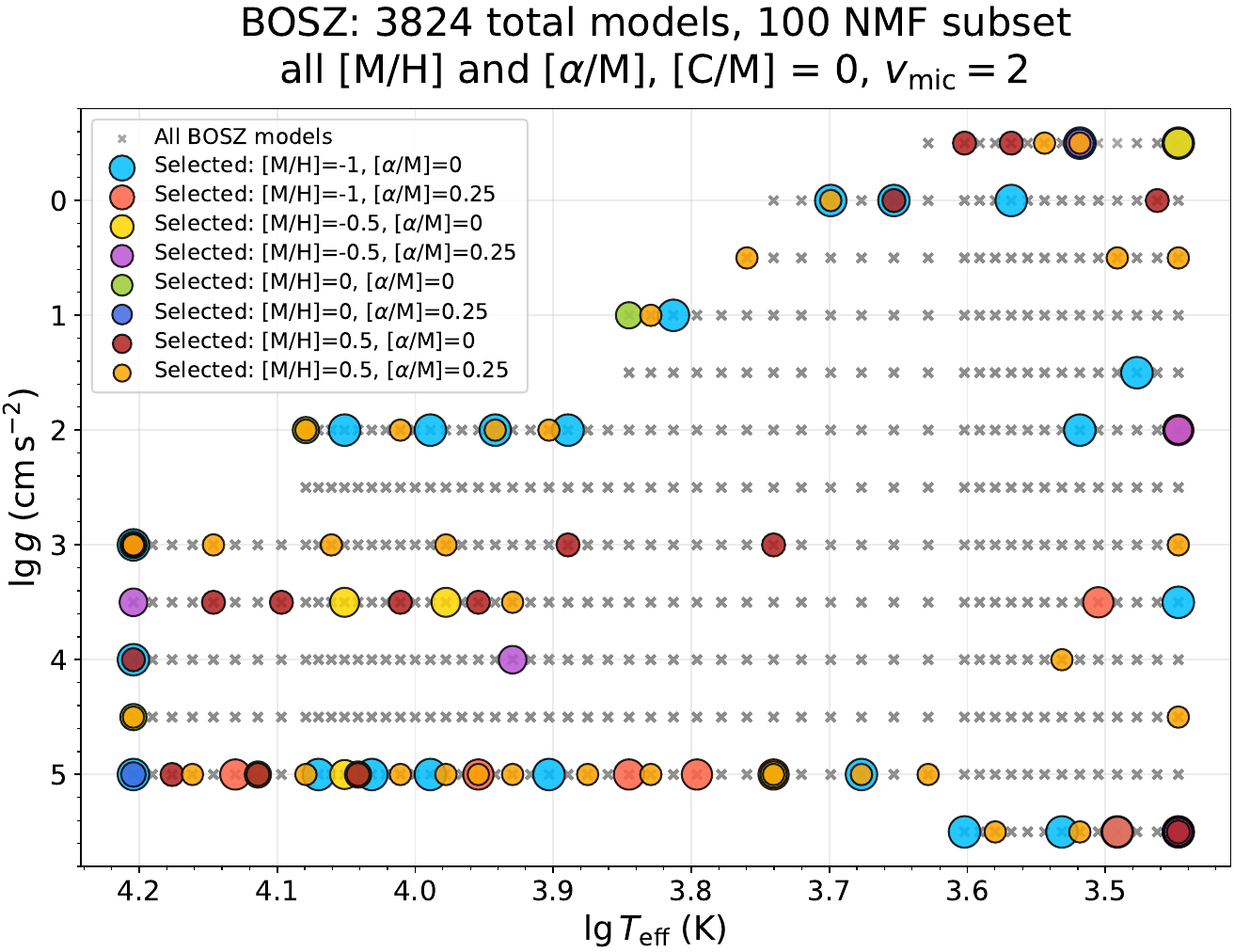}
    \includegraphics[width=.49\textwidth]{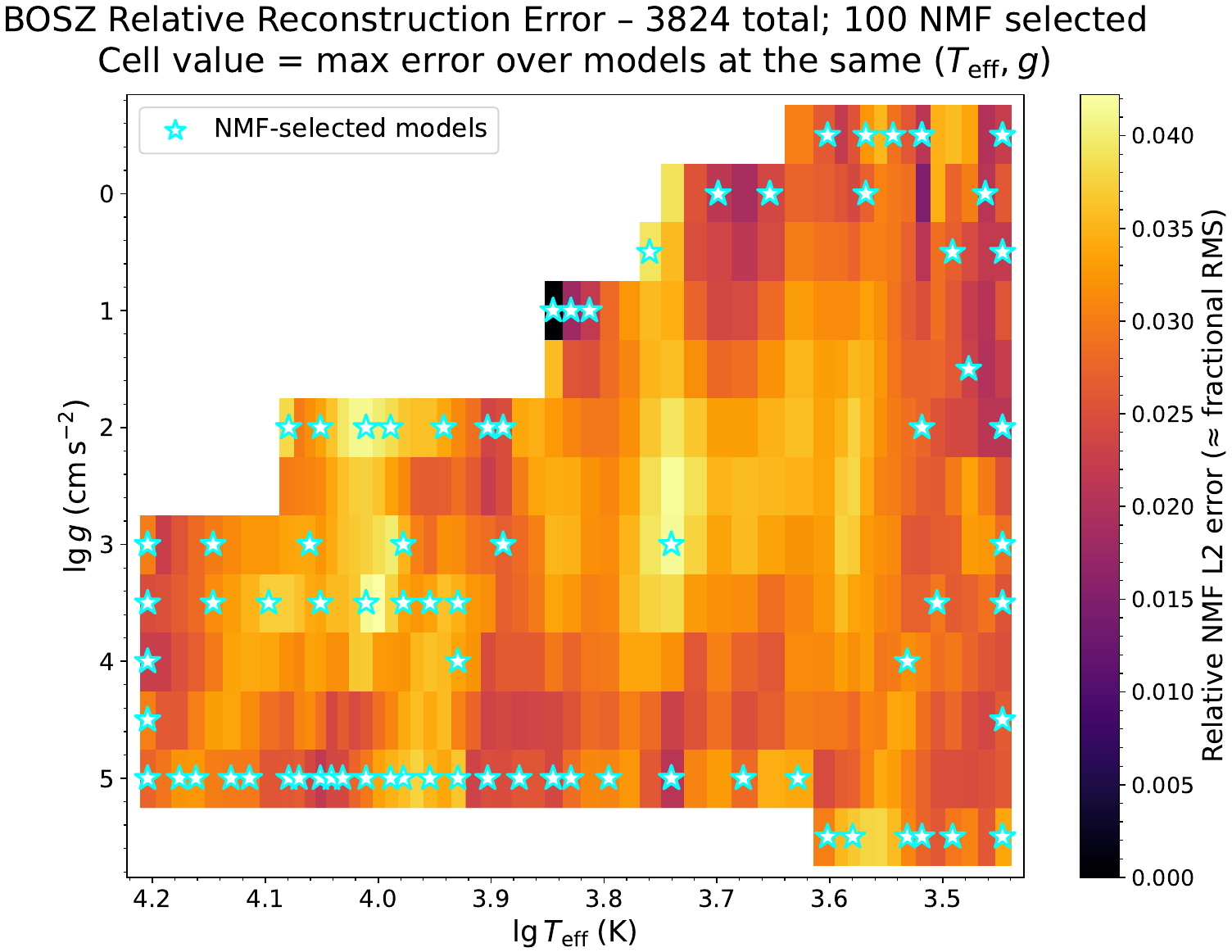}
    \caption{\textbf{Optimal template selection for the full multi-dimensional BOSZ library.} We applied the SNPA algorithm to the expanded BOSZ grid, allowing for variations in both metallicity ($-1.0 \le [\mathrm{M/H}] \le +0.5$) and $\alpha$-abundance ($[\alpha/\mathrm{M}] \in \{0, 0.25\}$), selecting a final subset of 100 optimal spectra for our \textsc{pPXF} kinematic extractions. \textbf{Left:} The distribution of the 100 selected models (large colored circles) superimposed on the full available grid (grey crosses). The colors correspond to different chemical abundance combinations as indicated in the legend. Consistent with the restricted test in \autoref{fig:nmf_demo}, the algorithm efficiently maps the multidimensional edges of the parameter space to maximize the additive spanning power of the subset. \textbf{Right:} The corresponding relative non-negative least squares (NNLS) RMS reconstruction error projected onto the $\log g$--$T_\mathrm{eff}$ plane (cyan circles denote the selected models). The reconstruction error is kept predominantly below $\sim 2\%$ everywhere, demonstrating that this highly compressed, computationally efficient 100-star subset perfectly captures the spectral variance of the massive parent library.}
    \label{fig:nmf_full}
\end{figure*}

Accurate kinematic extraction via full-spectrum fitting relies fundamentally on a stellar template library that can faithfully reproduce the absorption features of the target galaxy. In the optical regime, this is routinely achieved using extensive empirical stellar libraries such as MILES \citep{SanchezBlazquez2006, FalconBarroso2011} or the SDSS-IV MaStar library \citep{Yan2019}. However, these empirical libraries are generally truncated at $\lambda \approx 1\,\mu\mathrm{m}$, rendering them unsuitable for our JWST NIRSpec \texttt{G235H} observations ($1.66$--$3.17\,\mu\mathrm{m}$). While some ground-based empirical libraries have been extended into the near-infrared---such as the X-shooter Spectral Library \citep[XSL;][]{Verro2022}---they suffer from a critical limitation: ground-based spectra are interrupted by severe atmospheric telluric absorption bands. Utilizing these ground-based libraries would forfeit the unprecedented continuous wavelength coverage provided by JWST. 

To exploit the full, telluric-free spectral range of JWST, we must turn to purely synthetic spectral models. The use of synthetic stellar spectra for the extraction of galaxy stellar kinematics is exceptionally well established. Synthetic templates have been employed with \textsc{pPXF} since at least \citet{Cappellari2009apjl}, who used the SYNTHE models of \citet{Munari2005} as primary templates to measure galaxy velocity dispersions in the restframe UV at high redshift, an approach subsequently validated with early X-shooter observations \citep{vandeSande2011}. Over the following decade, high-resolution PHOENIX synthetic spectra\footnote{\url{https://phoenix.astro.physik.uni-goettingen.de/}} \citep{Husser2013} became a standard tool for near-infrared stellar kinematics, successfully cross-checked against empirical libraries in demanding dynamical black hole and nuclear star cluster measurements \citep{Seth2014, Nguyen2018, Nguyen2019}, and used as primary templates for optical Calcium Triplet kinematic extractions \citep{Thater2022}. More recently, the C3K synthetic stellar spectral library\footnote{To our knowledge, the high resolution version of the C3K library is not publicly available online.} developed by Conroy and collaborators \citep{Conroy2018} has been successfully employed to recover higher-order Gauss-Hermite moments \citep{DEugenio2023} and to map spatially resolved JWST stellar kinematics \citep{DEugenio2024Nature}. 

For JWST/NIRSpec observations in particular, a rapid succession of recent studies has firmly established PHOENIX synthetic spectra as a powerful tool for mapping stellar kinematics and dynamically measuring supermassive black holes in nearby galaxies and ultra-compact dwarfs \citep[e.g.,][]{Dumont2025, Tahmasebzadeh2025, Taylor2025, Nguyen2025, Nguyen2026maser, Nguyen2026m81}. By directly comparing synthetic grids with empirical libraries like XSL, these authors demonstrated that both yield highly consistent stellar kinematics (within $\sim 3-5\%$). Crucially, they explicitly highlighted that synthetic libraries possess the fundamental advantage of continuous, gap-free wavelength coverage, which is essential to fully exploit the pristine spectra delivered by JWST. However, applying synthetic spectra to JWST data is not without its challenges. \citet{AlAmri2026} recently evaluated PHOENIX against empirical libraries for JWST/NIRSpec observations of M87, finding that while the overall kinematics are robust, PHOENIX struggles to accurately reproduce certain specific absorption features (e.g., individual Ca~I lines), necessitating custom masking.

This highlights the need to continuously evaluate the latest generations of theoretical models against pristine JWST data. While \citet{Hill2022} successfully employed an earlier iteration of the BOSZ library to extract stellar parameters for the MaStar empirical library using \textsc{pPXF}, the significantly updated 2024 release\footnote{\url{https://archive.stsci.edu/hlsp/bosz}} \citep{Meszaros2024} has not yet been utilized as a kinematic template basis for integrated galaxy light. To assess its performance, we extracted the kinematics of our target galaxies under identical conditions using both the PHOENIX and the new BOSZ libraries. We found that the two grids yield highly consistent absolute velocity dispersions. However, the BOSZ models exhibited significantly fewer unphysical wiggles in the continuum and provided a noticeably superior match to the detailed depths of complex molecular blends. We therefore adopted the updated BOSZ library for all our dynamical analyses.

Our main methodological innovation in this section is not the use of synthetic spectra themselves, but rather the introduction of a mathematically optimal way to manage them. The price of modern synthetic libraries is the enormous volume and redundancy of the spectra; the BOSZ library contains hundreds of thousands of models. Evaluating highly correlated matrices of this size in \textsc{pPXF} is computationally prohibitive. Previous approaches to compress these libraries included manually selecting a few representative templates, drawing a uniform grid of parameters, or using hierarchical clustering \citep[as adopted in the MaNGA Data Analysis Pipeline;][]{Westfall2019}. 

However, neither uniform grids nor hierarchical clustering are mathematically optimal for full-spectrum fitting. Because \textsc{pPXF} fits the galaxy spectrum as a strictly \emph{non-negative} linear combination of stellar templates, the ideal subset of templates is the one that identifies the mathematical ``convex hull'' of the library's parameter space. As introduced in \autoref{sec:nmf_method}, this is precisely achieved by near-separable Non-Negative Matrix Factorization (NMF).

To visualize the mechanics of this algorithm, we first ran a restricted demonstration. We limited the BOSZ library strictly to solar metallicity ($[\mathrm{M/H}]=0$) and solar $\alpha$-abundance ($[\alpha/\mathrm{M}]=0$), and tasked the SNPA algorithm to extract an optimal subset of 20 spectra. The results, shown in \autoref{fig:nmf_demo}, are initially counterintuitive: instead of selecting models that uniformly tile the $\log g - T_{\rm eff}$ plane, the NMF algorithm aggressively selects spectra that lie almost exclusively on the extreme boundaries and corners of the parameter space. Remarkably, despite having no prior knowledge of the physical parameters, the selected templates (the star symbols in \autoref{fig:nmf_demo}) not only identify the edges of the parameter ranges, but even pinpoint the specific corners of the jagged distribution. Mathematically, this is the exact required behavior. Any interior spectrum can be perfectly reproduced by a non-negative linear combination of the boundary spectra that bracket it, whereas an extreme boundary spectrum cannot be synthesized by mixing interior spectra without employing unphysical negative weights. Thus, the NMF algorithm naturally identifies the absolute minimum basis set required to span the spectral variance of the library.

For our final kinematic extraction, we deployed the NMF algorithm across the fully expanded BOSZ grid, allowing for variations in both $[\mathrm{M/H}]$ and $[\alpha/\mathrm{M}]$. We extracted an optimal, highly compressed subset of 100 actual synthetic spectra. \autoref{fig:nmf_full} illustrates the distribution of these selected models across all four physical parameters. Despite the massive reduction in the number of templates, the relative non-negative least squares (NNLS) reconstruction error across the entire parent library is kept overwhelmingly below $\sim 2\%$. 

This highly compressed 100-star subset perfectly encapsulates the full additive spanning power of the complete BOSZ library, ensuring maximum accuracy in our kinematic extractions at a fraction of the computational cost. Furthermore, as we will demonstrate in the following sections, these NMF-selected BOSZ spectra provide remarkably accurate fits to the high-quality JWST observations, proving that modern synthetic models are now more than capable of reproducing real stellar populations accurately enough for precision dynamics.

\section{Stellar Kinematic Extraction with \textsc{pPXF}}
\label{sec:kinematics}

\begin{figure*}
    \centering
    \includegraphics[width=\textwidth]{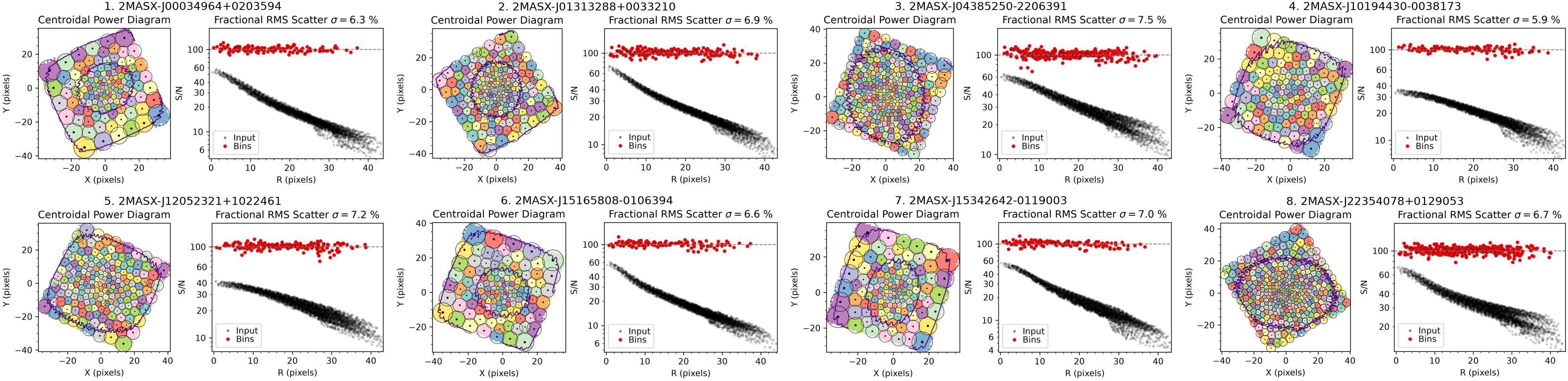}
    \caption{\textbf{Spatial binning and fit residuals.} \textit{Left sub-panels:} The spatial binning maps for the galaxies in our sample, generated using the \textsc{PowerBin} algorithm to achieve a target $\mathrm{S/N} = 100$ per bin. The algorithm produces highly compact, geometrically regular bins that optimally preserve spatial resolution in the galaxy cores while building sufficient $\mathrm{S/N}$ in the outer halos. \textit{Right sub-panels:} The fractional RMS scatter of the \textsc{pPXF} residuals as a function of the bin radius. The red dots represent the binned spectra, while the grey dots show the raw input spaxels. The robust scale-invariant clipping implemented in \textsc{pPXF} v9.5 ensures consistent and low-scatter residuals across the entire field of view.}
    \label{fig:kin_powerbin}
\end{figure*}

\begin{figure*}
    \centering
    \includegraphics[width=\textwidth]{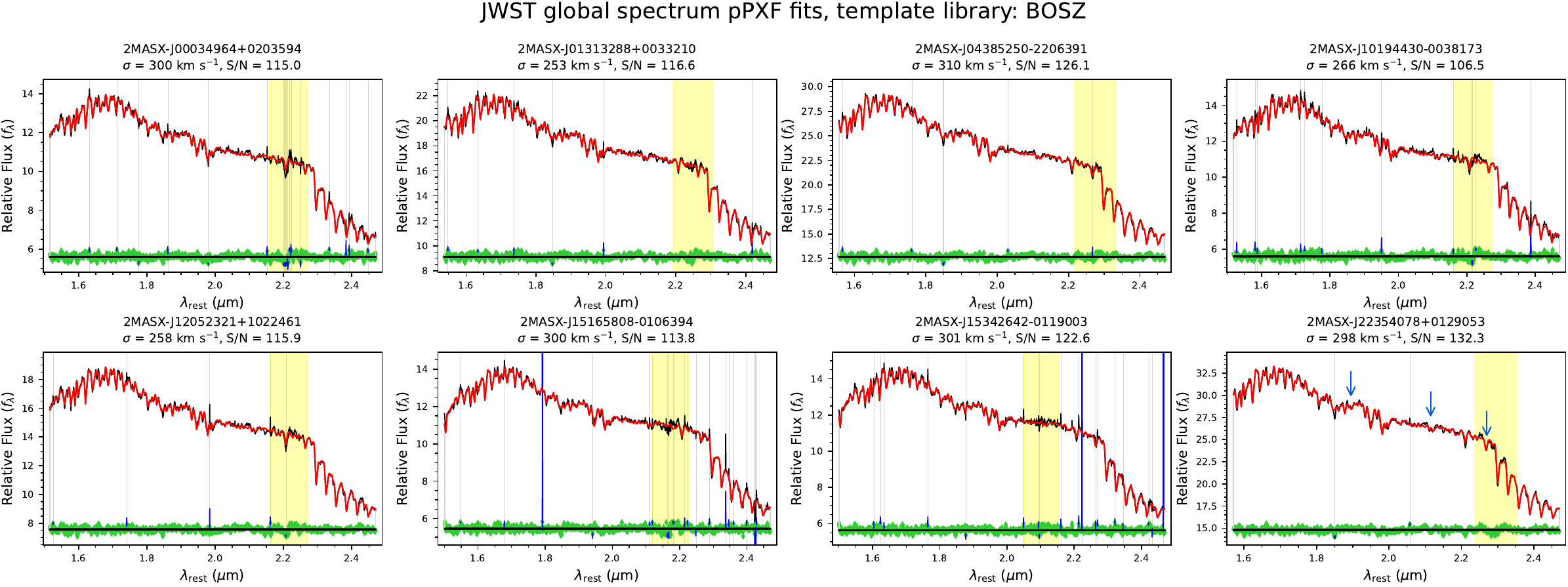}
    \caption{\textbf{\textsc{pPXF} fits to globally integrated spectra using BOSZ templates.} Each panel displays the fit to a high-$\mathrm{S/N}$ spectrum integrated over all spaxels. To account for the spatially varying NIRSpec detector gap, we co-add only the unaffected spaxels at each wavelength and scale the result to correct for the missing flux. The observed JWST/NIRSpec spectra are shown in black, the best-fitting combinations of NMF-selected BOSZ templates in red, and the fit residuals in green. Grey bands mark regions excluded from the \textsc{pPXF} fit (with their residuals plotted in blue). Light-yellow bands highlight wavelengths where the detector gap crosses the field of view; the reduced number of contributing spaxels in these regions degrades the spectrum's quality and can introduce mild continuum artifacts. In the 2MASX J22354078+0129053 panel, vertical blue arrows mark the same three wavelengths highlighted in the PHOENIX comparison (\autoref{fig:kin_global_phoenix}), pointing out the most prominent localized differences between the two models. Beyond these specific features, the BOSZ fits generally yield lower RMS residuals, achieving a median S/N across these panels that is 24\% higher than PHOENIX (as discussed in the text).}
    \label{fig:kin_global_bosz}
\end{figure*}

\begin{figure*}
    \centering
    \includegraphics[width=\textwidth]{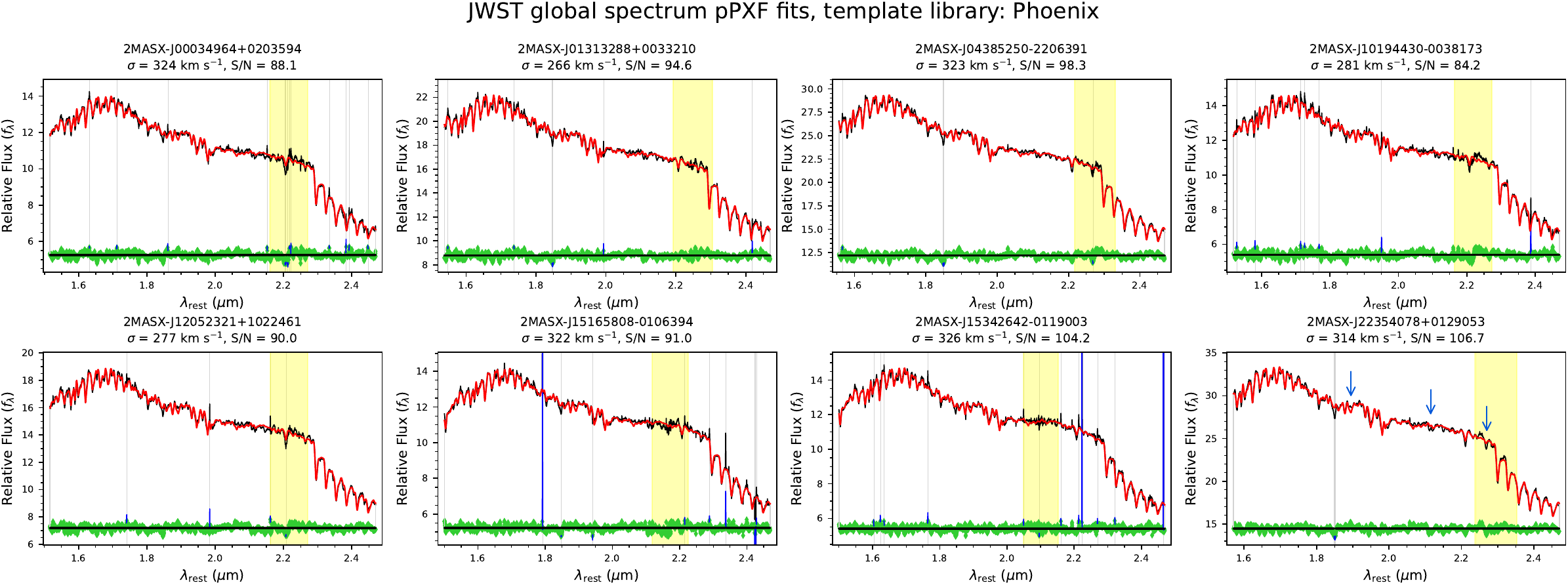}
    \caption{\textbf{\textsc{pPXF} fits to the globally integrated spectra using PHOENIX templates.} The layout, global spectra, and shaded regions are the same as in \autoref{fig:kin_global_bosz}, but the stellar templates are drawn from the PHOENIX library. The three blue arrows in the panel for 2MASX J22354078+0129053 identify the main systematic residuals for which BOSZ provides a better match. The rightmost arrow marks the Ca~I absorption feature that is missing from PHOENIX, as also noted by \citet{AlAmri2026}.}
    \label{fig:kin_global_phoenix}
\end{figure*}

\begin{figure*}
    \centering
    \includegraphics[width=\textwidth]{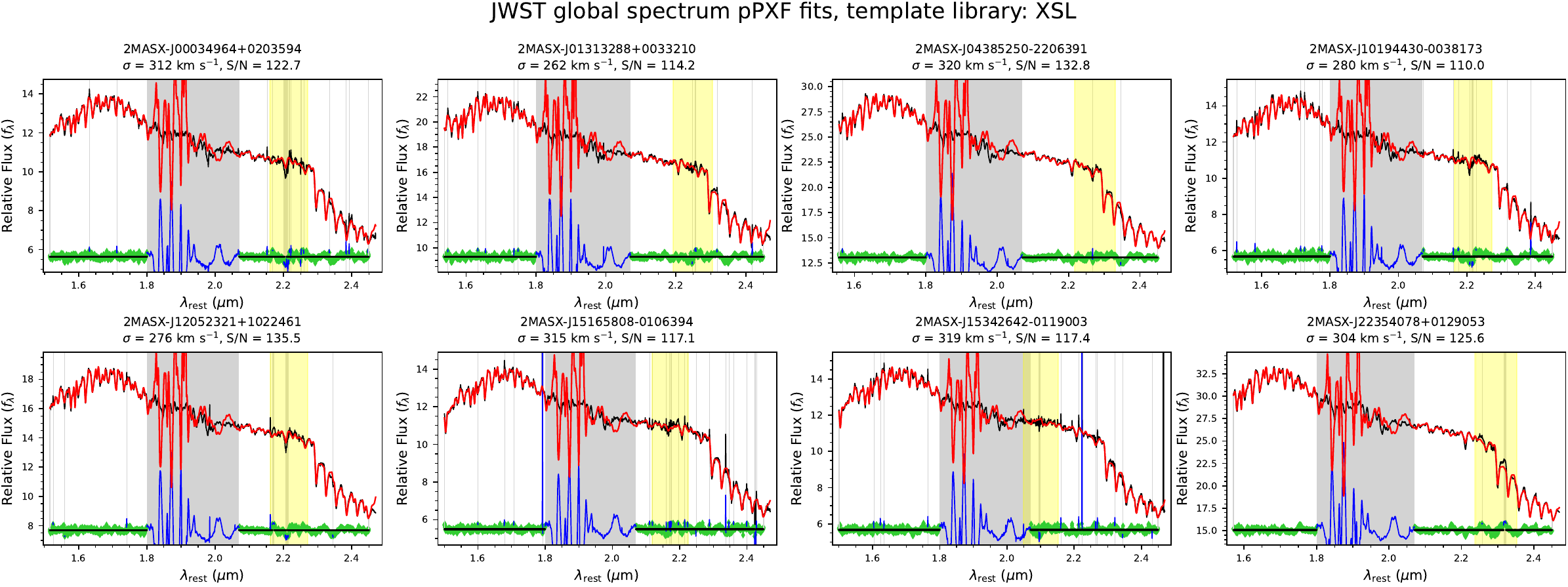}
    \caption{\textbf{\textsc{pPXF} fits to the globally integrated spectra using XSL templates.} The layout, global spectra, and light-yellow detector-gap regions are the same as in \autoref{fig:kin_global_bosz}, but the stellar templates are drawn from the empirical X-shooter Spectral Library. The broad grey band is excluded from the \textsc{pPXF} fit because of the ground-based telluric gap in XSL. In the wavelength region available in common with BOSZ, XSL does not provide an appreciably better match, while its large gaps make it impractical for exploiting the continuous JWST wavelength coverage.}
    \label{fig:kin_global_xsl}
\end{figure*}

\begin{figure*}
    \centering
    \includegraphics[width=\textwidth]{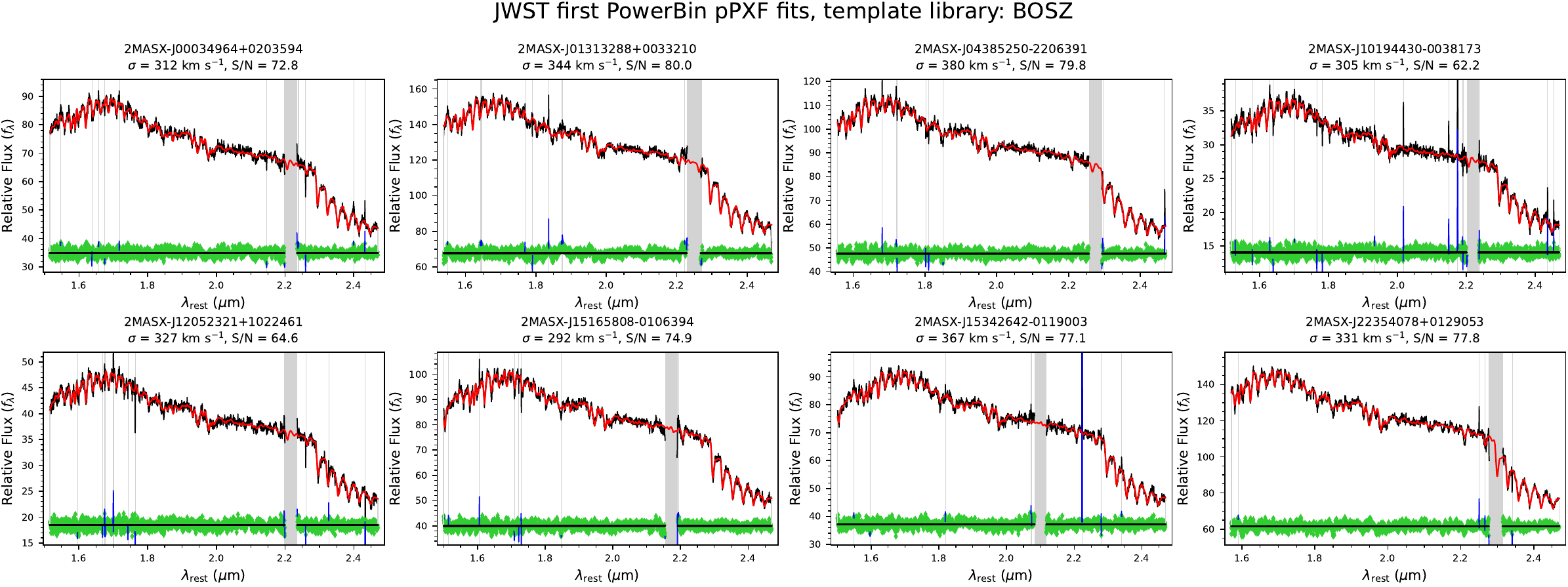}
\caption{\textbf{\textsc{pPXF} kinematic fits to the central spatial bins.} Similar to \autoref{fig:kin_global_bosz}, but showing the fit to the central-most spatial bin (Power bin 1) for each galaxy. These spectra probe the kinematics deep within the gravitational sphere of influence of the supermassive black holes. Despite the extreme intrinsic velocity dispersions ($\sigma \gtrsim 300\,\mathrm{km\,s^{-1}}$), the BOSZ synthetic templates successfully track the significantly broadened absorption features, particularly the deep $^{12}$CO bandheads longward of $2.29\,\mu\mathrm{m}$. Note that the quoted $\mathrm{S/N}$ in each panel is empirically computed from the fits as the median of the spectrum divided by the biweight dispersion of the fit residuals, rather than from the formal pipeline-propagated spectral variance. Consequently, this quoted value represents a lower limit to the formal spectral $\mathrm{S/N}$, as it encompasses both true random noise and any systematic effects or fit inaccuracies. For this reason, the $\mathrm{S/N}$ is not identical across figures showing different template fits to the same data; instead, a higher $\mathrm{S/N}$ directly indicates a higher-quality fit.}
    \label{fig:kin_central}
\end{figure*}

\begin{figure*}
    \centering
    \includegraphics[width=\textwidth]{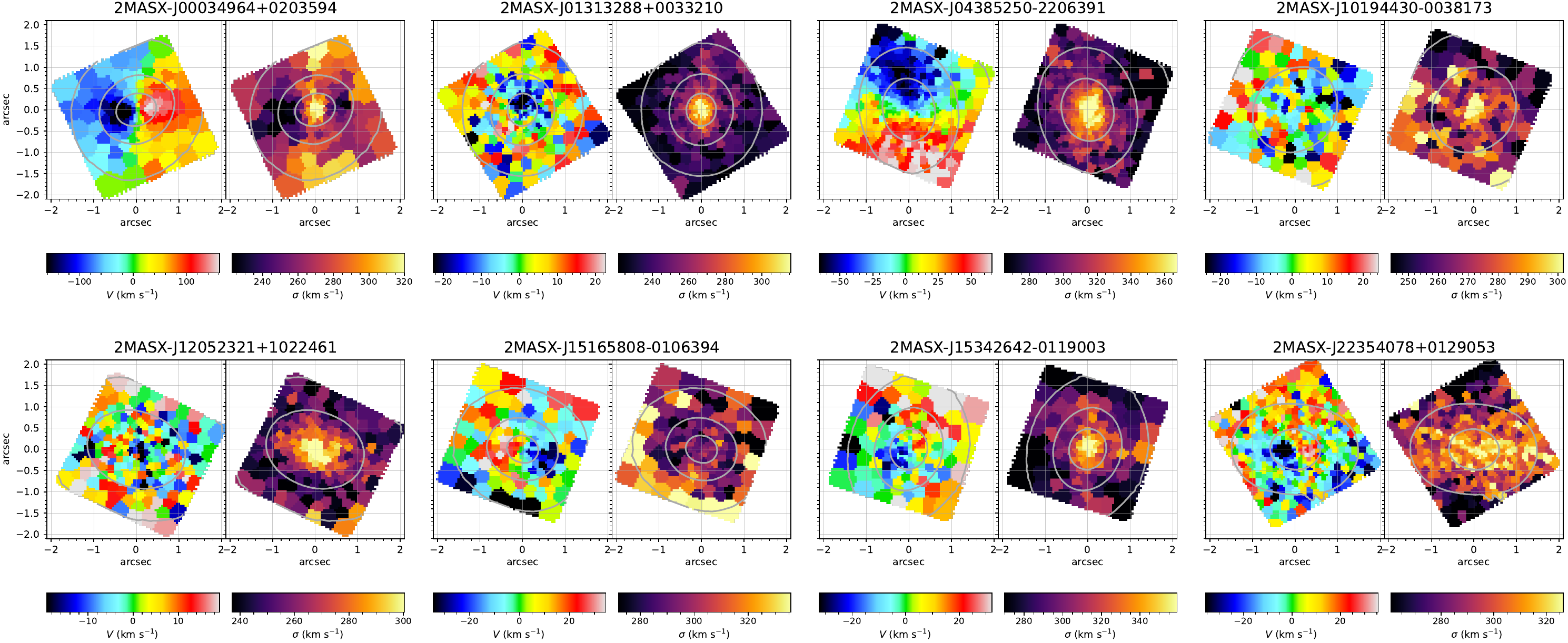}
    \caption{\textbf{Stellar kinematic maps derived from JWST/NIRSpec.} For each galaxy, the left panel displays the mean stellar velocity field ($V$), and the right panel displays the stellar velocity dispersion field ($\sigma$). The maps were extracted using \textsc{pPXF} with the optimal 100-star subset of the BOSZ synthetic library. Thanks to the exquisite spatial resolution and stability of the JWST PSF, a clear dynamical distinction stands out among the sample. Most of the galaxies exhibit a steep, abrupt central peak in velocity dispersion, which is the hallmark dynamical signature of a central ultramassive black hole. In stark contrast, galaxies 5 and 7 deviate from this pattern: galaxy 5 shows a central drop or a constant $\sigma$ plateau, while galaxy 7 exhibits only a marginal, smooth central rise rather than the sharp Keplerian spike seen in the rest of the sample.}
    \label{fig:kin_maps}
\end{figure*}

With the optimal synthetic template library selected, we turn to the extraction of the spatially resolved stellar kinematics from our JWST NIRSpec IFU observations. The spectral fitting was performed using the \textsc{pPXF} software, which operates in pixel space to recover the line-of-sight velocity distribution (LOSVD) while simultaneously fitting for the stellar continuum.

\subsection{The Telluric Gap Challenge and the Choice of Templates}

In our initial iterations of the kinematic extraction, we utilized the empirical X-shooter Spectral Library \citep[XSL;][]{Verro2022}. Because XSL is constructed from ground-based observations, it is fundamentally affected by the opacity of the Earth's atmosphere. The most severe telluric absorption gap between the $H$ and $K$ bands falls squarely within the rest-frame wavelength coverage of our NIRSpec \texttt{G235H} grating.

Although the XSL Data Release 3 \citep{Verro2022} reports the primary ``NIR c'' telluric region as spanning $1.81$--$1.93\,\mu\mathrm{m}$, we found in practice that forcing \textsc{pPXF} to fit empirical XSL templates outside this strict window yielded unacceptable continuum residuals and strong template-mismatch artifacts. To obtain stable fits with XSL, we were forced to aggressively mask the rest-frame wavelength range from $1.80\,\mu\mathrm{m}$ to $2.07\,\mu\mathrm{m}$. This massive masking effectively excised a major continuous segment of our JWST data, discarding valuable kinematic information and defeating the purpose of space-based, continuous near-infrared spectroscopy.

To make this comparison explicit, we show otherwise consistent global-spectrum fits using XSL, PHOENIX, and BOSZ in \autoref{fig:kin_global_xsl}, \autoref{fig:kin_global_phoenix}, and \autoref{fig:kin_global_bosz}, respectively. The figures provide the direct visual basis for the template choice discussed above: the large gaps required for XSL make it impractical, while PHOENIX leaves some individual features less well reproduced than BOSZ. The three arrows in the panel for 2MASX J22354078+0129053 highlight the main systematic differences, with the rightmost arrow marking the Ca~I feature missing from PHOENIX that was previously identified by \citet{AlAmri2026}. More generally, the $\mathrm{S/N}$ reported in each panel is the median signal divided by the standard deviation of the fit residuals. Because the signal is identical in the BOSZ and PHOENIX fits, their $\mathrm{S/N}$ ratio directly measures the inverse ratio of the residual scatter. BOSZ gives a higher $\mathrm{S/N}$ in all seven galaxies, with a median pairwise increase of $24\%$, corresponding to a median reduction of $19\%$ in the residual standard deviation. In the wavelength region available in common, XSL does not provide an appreciably better fit than BOSZ. For all final kinematic extractions presented below, we exclusively utilized the optimal, NMF-selected 100-star subset of the BOSZ library.

\subsection{Extraction Procedure, \textsc{PowerBin} and \textsc{pPXF} Configuration}

Our fitting procedure was strictly standardized across the entire ultramassive galaxy sample. We restricted the \textsc{pPXF} fit to the rest-frame wavelength range $1.50$--$2.47\,\mu\mathrm{m}$, with the upper limit also motivated by our comparison with the XSL library, which does not extend to longer wavelengths. This range is exceptionally rich in stellar absorption features, notably the deep $^{12}$CO vibrational bandheads longward of $2.29\,\mu\mathrm{m}$, as well as numerous CN and OH molecular lines. At wavelengths beyond approximately $2.5\,\mu\mathrm{m}$, while within our grating range, the spectra contain comparatively few additional features useful for constraining the stellar kinematics.

Before applying \textsc{PowerBin}, we excluded spaxels with insufficient accumulated exposure, identified using the cube flags and caused by the four offset dither positions. This slightly reduced the field of view, but avoided the formation of large outer PowerBin bins needed to compensate for the sharp decline in surface brightness. We found this preferable to formally including all spaxels, which produced outer bins dominated by systematic effects. Specifically, we summed the \texttt{WMAP} values over wavelength and retained only spaxels with a total exposure count exceeding $80\%$ of the median positive value. We then spatially binned the remaining NIRSpec spaxels using the Centroidal Power Diagram binning algorithm and software\footnote{\url{https://pypi.org/project/powerbin/}} (\textsc{PowerBin}, \citealt{Cappellari2025}) to achieve the signal-to-noise ratio ($\mathrm{S/N}$) required for robust extraction of the velocity ($V$) and velocity dispersion ($\sigma$). We set a target $\mathrm{S/N} = 100$ per bin, evaluated across the continuum. Unlike traditional Voronoi binning, \textsc{PowerBin} produces exceptionally smooth and geometrically compact bins even in the presence of complex detector footprints, ensuring optimal spatial resolution in the high-surface-brightness galaxy cores while reliably aggregating flux in the faint outer regions (\autoref{fig:kin_powerbin}).

A critical instrumental feature of the NIRSpec IFU is the physical gap between the two detector arrays, which creates an unobserved wavelength band in the observed frame (typically between $2.36$ and $2.49\,\mu\mathrm{m}$ for the \texttt{G235H} grating). Rather than applying a global mask that would penalize the entire field of view, we configured \textsc{pPXF} to dynamically mask the detector gap on a strictly per-spaxel basis, relying on the local weight map (\texttt{WMAP}) to identify and exclude pixels with zero coverage. Because the wavelength of the gap varies spatially across the IFU field of view, the global spectra in \autoref{fig:kin_global_bosz}, \autoref{fig:kin_global_phoenix}, and \autoref{fig:kin_global_xsl} were constructed wavelength by wavelength by co-adding only the spaxels whose corresponding pixel does not fall within the gap. We then corrected the co-added spectrum for the estimated flux of the excluded spaxels. The light-yellow bands in the figures mark wavelengths affected by this procedure: they are not excluded from the fit, but their lower effective spatial coverage explains the reduced spectral quality and mild continuum artifacts within those intervals.

For each spatial bin, \textsc{pPXF} solved for the first two kinematic moments ($V$ and $\sigma$). To account for differences in continuum shape between the BOSZ synthetic templates and the JWST data (arising from dust attenuation and flux calibration uncertainties), we included an 8th-degree additive polynomial, to allow for possible variations in the line strength of the population, and a 1st-degree multiplicative polynomial, to account for possible calibration inaccuracies. We note that the derived kinematics are generally weakly sensitive to the adopted polynomial degrees, yielding qualitatively very similar results across a relatively large range of values. While it is common practice in full-spectrum fitting to rely exclusively on additive polynomials, we have opted to include a multiplicative component as well. The multiplicative polynomial can be kept at a much lower degree than the additive one while producing nearly indistinguishable results, as long as both are present to some degree. For instance, increasing our multiplicative degree from 1 to 8 to match the additive component would make no difference to the final fit, other than resulting in a longer computation time. This computational penalty arises because multiplicative polynomials must be optimized non-linearly, whereas additive polynomials are solved highly efficiently via linear optimization.

\subsection{Robust Fitting: Improvements in \textsc{pPXF} v9.5}

The presence of residual bad pixels, cosmic rays, and detector artifacts in IFS cubes necessitates robust outlier rejection during the spectral fitting process. For this work, we utilized the most recent version of the software\footnote{\url{https://pypi.org/project/ppxf/}} (\textsc{pPXF} v9.5), which introduces major improvements to the robustness and speed of the fits when utilizing the \texttt{clean} keyword. 

In previous iterations of the software, sigma-clipping required an accurately normalized \texttt{noise} spectrum to function correctly and, in most situations, also required a separate preliminary \textsc{pPXF} call to estimate the residuals before invoking the clipping call. In v9.5, the clipping algorithm has been overhauled to be scale-invariant. Consequently, setting \texttt{clean=True} now performs a fast, robust, and mathematically rigorous outlier rejection that is completely independent of the absolute scaling of the input noise array. The clipping threshold also adapts to the number of currently fitted spectral pixels, $N_{\rm good}$. Specifically, $\texttt{clip}=-\Phi^{-1}(0.5/N_{\rm good})$, where $\Phi$ is the standard normal cumulative distribution function, is the number of standard deviations used for rejecting outliers, rather than a fixed value such as the commonly used $\texttt{clip}=3$. This approximately allows one false Gaussian outlier on average, independently of the spectrum length, while the clipping is iterated until convergence. Given its stability and efficiency, the use of the \texttt{clean=True} keyword is now strongly recommended for most realistic observational situations. We utilized this scale-invariant clipping across all our bins, ensuring that our derived velocity dispersions are completely uncontaminated by residual detector artifacts (\autoref{fig:kin_global_bosz} to \autoref{fig:kin_central}).

The final, spatially resolved kinematic maps for the galaxy sample are presented in \autoref{fig:kin_maps}. The exceptionally high $\mathrm{S/N}$ of the JWST data, combined with the continuous wavelength coverage of the NMF-selected BOSZ templates and the rigorous \textsc{pPXF} outlier rejection, yields exquisitely smooth and dynamically revealing velocity and velocity dispersion fields.

\subsection{Overview of the Extracted Stellar Kinematics}
\label{sec:kinematic_overview}

While the primary goal of this kinematic extraction is to provide robust, high-$\mathrm{S/N}$ inputs for our dynamical black hole mass measurements, the exquisite velocity and velocity dispersion fields (\autoref{fig:kin_maps}) also offer valuable insights into the central orbital structure of these ultramassive galaxies. 

It is important to emphasize that the $3\arcsec \times 3\arcsec$ field of view of the NIRSpec IFU typically maps only the inner few kiloparsecs of these giant ellipticals, falling well short of one effective radius ($1\,R_\mathrm{e}$). Consequently, the spatial coverage is insufficient to assign a definitive global kinematic classification (e.g., via the $\lambda_R$ parameter integrated over the half-light radius). Nevertheless, the central kinematics of our sample are overwhelmingly consistent with what is expected for massive slow rotators \citep{Emsellem2011p3}. Fast rotators almost never exhibit such a stark lack of ordered rotation in their innermost regions \citep{Krajnovic2011p2}. See reviews in \citet{Cappellari2016,Cappellari2026} for a comprehensive discussion of the kinematic classification of early-type galaxies.

For example, 2MASX-J10194430-0038173 and 2MASX-J12052321+1022461 show velocity fields with virtually no ordered rotation at all, firmly placing them in the slow-rotator category. Another common dynamical feature in massive early-type galaxies is the presence of a Kinematically Decoupled Core (KDC). This is prominently visible in 2MASX-J00034964+0203594, where a clear central rotation signature peaks and then begins dropping back toward zero velocity at the very edges of the NIRSpec field of view. Similar, albeit slightly less pronounced, KDCs are also present in the velocity fields of 2MASX-J15165808-0106394 and 2MASX-J15342642-0119003, and one is exceedingly obvious in the center of 2MASX-J22354078+0129053.

The only possible exception to the ubiquitous slow-rotator nature of this sample is 2MASX-J04385250-2206391. Its velocity field displays a coherent rotation pattern that continues to rise at the outer boundary of the NIRSpec field. Based on the current limited spatial coverage, it is impossible to determine whether this object is genuinely a fast rotator, or if we are simply witnessing the inner regions of a KDC that is physically larger than our $3\arcsec$ field of view.

Finally, the kinematics of 2MASX-J22354078+0129053 warrant special mention. In addition to its prominent inner KDC, its velocity dispersion map exhibits a striking elongation aligned with the galaxy's photometric major axis. This morphology is the classic signature of the so-called ``$2\sigma$'' galaxies \citep{Krajnovic2011p2}, which are typically characterized by two counter-rotating disc components. However, an identical kinematic signature can naturally arise in massive slow rotators that are weakly triaxial but dynamically dominated by families of counter-rotating tube orbits (the same orbital families that dominate axisymmetric systems). With the caveat that our field of view only covers the central region, the NIRSpec kinematics for 2MASX-J22354078+0129053 are remarkably consistent with well-studied giant ellipticals such as NGC\,4365 and NGC\,5813. Detailed Schwarzschild orbit-superposition models of those galaxies \citep{vandenBosch2008, Krajnovic2015} concluded that they are indeed dominated by a population of counter-rotating tube orbits, which simultaneously generate both the central KDC and the extended major-axis velocity dispersion peaks observed here.

\section{Conclusions}
\label{sec:conclusions}

We have presented the foundational data products for the JWST Ultramassive Galaxy Sample, a targeted observational program designed to dynamically measure the central supermassive black holes in a representative subset of 8 galaxies drawn from a full-sky parent census of 101 extreme early-type systems ($M_\star \gtrsim 2\times 10^{12}\,\mathrm{M}_\odot$). By targeting this exceptional mass regime across a diverse range of galactic environments, this project ultimately aims to determine whether black hole scaling relations undergo a fundamental transition at the top of the galaxy mass hierarchy, shifting from being driven by velocity dispersion ($\sigma_\star$) to total stellar mass ($M_\star$). 

Because robust dynamical black hole mass measurements are strictly bottlenecked by the quality of the photometric mass models and the fidelity of the stellar kinematics, this first paper has focused entirely on optimizing these critical data extraction processes. Our main methodological contributions and observational findings are summarized as follows:

\begin{enumerate}
    \item \textbf{Multi-scale Photometric Modelling:} To properly constrain the gravitational potential from the black hole sphere of influence out to the extended dark matter halo, we seamlessly combined diffraction-limited JWST imaging (NIRCam and collapsed NIRSpec cubes) with wide-field ground-based $z$-band imaging from the DESI Legacy Surveys. Using the Multi-Gaussian Expansion (MGE) method, we simultaneously fitted these matched datasets to produce high-fidelity surface brightness models that accurately track the galaxies across several orders of magnitude in spatial scale.
    
    \item \textbf{Optimal Template Selection via NMF:} To fully exploit the continuous, telluric-free $1.66$--$3.17\,\mu\mathrm{m}$ wavelength coverage of the JWST NIRSpec \texttt{G235H} grating, we moved beyond ground-based empirical spectral libraries, which suffer from severe atmospheric absorption gaps. Instead, we adopted the updated BOSZ synthetic stellar library. To overcome the computational prohibitive size and massive redundancy of modern synthetic grids, we pioneered the use of near-separable Non-Negative Matrix Factorization (NMF). Using the SNPA algorithm, we isolated an optimal, highly compressed subset of 100 synthetic spectra that form the mathematical convex hull of the library. This subset perfectly captures the additive spanning power of the full library, establishing a highly efficient and robust new standard for full-spectrum fitting with JWST.
    
    \item \textbf{Robust Kinematic Extraction:} We extracted the spatially resolved stellar kinematics using \textsc{pPXF} v9.5. The combination of the \textsc{PowerBin} algorithm (which guarantees compact, high-$\mathrm{S/N}$ spatial bins) and the new scale-invariant \texttt{clean} keyword in \textsc{pPXF} ensured that our extracted velocity and velocity dispersion fields are exquisitely smooth and entirely free from residual detector artifacts or cosmic rays. 
    
    \item \textbf{Central Dynamical Signatures:} Despite the small field of view of the NIRSpec IFU, the kinematic maps unequivocally confirm that these ultramassive systems are dominated by slow-rotator dynamics. Kinematically Decoupled Cores (KDCs) are a nearly ubiquitous feature in the sample. Notably, we observed a classic $2\sigma$ major-axis velocity dispersion elongation in 2MASX-J22354078+0129053, a signature strongly consistent with a triaxial system dominated by counter-rotating tube orbits.
    
    \item \textbf{Signatures of Ultramassive Black Holes:} The exquisite spatial resolution of JWST clearly separates the sample into two distinct dynamical categories at the very center. The vast majority of the galaxies exhibit a steep, sharp central rise in their velocity dispersion fields---the hallmark Keplerian signature of a central ultramassive black hole. Conversely, two targets (2MASX-J12052321+1022461 and 2MASX-J22354078+0129053) show a flat or dropping central dispersion profile, suggesting either distinct orbital anisotropies, cored stellar density profiles, or differing black hole mass fractions.
\end{enumerate}

The customized techniques developed in this work---specifically the integration of JWST and DESI imaging via MGE, and the NMF-driven compression of synthetic spectral libraries for \textsc{pPXF}---provide a rigorous, reproducible framework for precision extragalactic astrophysics in the JWST era. In Paper II \citep{Cappellari2026p2} of this series, we utilize these optimized photometric models and kinematic maps as the foundational inputs for Jeans Anisotropic Modelling (JAM). This will allow us to dynamically weigh the central dark masses, reconstruct the orbital distributions, and definitively anchor the extreme upper boundary of the supermassive black hole scaling relations, which are presented in Paper~III \citep{Cappellari2026p3}.

%%%%%%%%%%%%%%%%%%%%%%%%%%%%%%%%%%%%%%%%%%
 
\section*{Acknowledgements}

This work is based on observations made with the NASA/ESA/CSA James Webb Space Telescope. The data were obtained from the Mikulski Archive for Space Telescopes at the Space Telescope Science Institute, which is operated by the Association of Universities for Research in Astronomy, Inc., under NASA contract NAS 5-03127 for JWST. These observations are associated with JWST programme 8217.   D.D.N. acknowledges support from the ELT postdoctoral research fellowship at the Department of Astronomy, University of Michigan.

This project used data products from the DESI Legacy Imaging Surveys \citep{Dey2019}, which consist of three individual and complementary projects: the Dark Energy Camera Legacy Survey (DECaLS), the Beijing-Arizona Sky Survey (BASS), and the Mayall $z$-band Legacy Survey (MzLS). Pipeline processing and analyses were supported by NSF's NOIRLab and the Lawrence Berkeley National Laboratory (LBNL). Legacy Surveys was supported by the Director, Office of Science, Office of High Energy Physics of the U.S. Department of Energy; the National Energy Research Scientific Computing Center; the U.S. National Science Foundation, Division of Astronomical Sciences; the National Astronomical Observatories of China, the Chinese Academy of Sciences and the Chinese National Natural Science Foundation. The complete acknowledgments can be found at \url{https://www.legacysurvey.org/acknowledgment/}. 

The authors used generative AI assistants to improve workflow efficiency, including language editing and programming support; all scientific ideas, analyses, interpretations, and conclusions are their own, and the authors take full responsibility for the content of this work.

\section*{Data Availability}
 
All observational data, extracted kinematics, and derived data products underlying this article will be made available as supplementary material alongside the published version.

\appendix

\label{lastpage}

\end{document}